\documentclass[letterpaper,11pt]{article}
\usepackage[margin=1in]{geometry}
\usepackage{amsthm, amsfonts,mathrsfs,amssymb,amsmath,clrscode,multirow,graphicx,amsmath,epic,eepic,url,bigstrut}
\usepackage{graphicx}
\usepackage{tikz}
\usepackage{pgfplots}

\newcommand{\F}{\mathbb{F}} 
\newcommand{\field}{\mathbb{F}} 

\newcommand{\N}{\mathbb N}
\newcommand{\Pp}{\mathcal P}
\newcommand{\Q}{\mathbb Q}
\newcommand{\R}{\mathbb R}
\newcommand{\degen}{\unlhd}

\newcommand{\braket}[1]{\langle #1 \rangle}
\newcommand{\ceil}[1]{\left\lceil #1 \right\rceil}

\newcommand{\sfloor}[1]{\lfloor #1 \rfloor}
\newcommand{\Ni}[2]{\ensuremath{[\![#1,#2]\!]}}

\newcommand{\calT}{\mathcal{T}}
\newcommand{\calN}{\mathcal{N}}
\newcommand{\calO}{\mathcal{O}}

\newcommand{\calM}{\mathcal{M}}
\newcommand{\Nn}{\mathcal{N}}

\newcommand{\Cc}{\mathscr{C}}

\newtheorem{theorem}{Theorem}[section]

\newtheorem{proposition}{Proposition}[section]
\newtheorem{definition}{Definition}[section]

\newenvironment{proof-sketch}{\trivlist\item[]\emph{Brief proof sketch}:}%
{\unskip\nobreak\hskip 1em plus 1fil\nobreak$\Box$
\parfillskip=0pt%
\endtrivlist}

\begin{document}
\title{Improved Rectangular Matrix Multiplication\\ using Powers of the Coppersmith-Winograd Tensor}
\author{Fran{\c c}ois Le Gall\\
Graduate School of Informatics\\
Kyoto University\\
\url{legall@i.kyoto-u.ac.jp}
\and
Florent Urrutia\\
IRIF\\
Universit\'e Paris Diderot\\
\url{urrutia@irif.fr}
}
\date{}

\maketitle
\thispagestyle{empty}
\setcounter{page}{0}
\begin{abstract}
In the past few years, successive improvements of the asymptotic complexity of square matrix multiplication have been obtained by developing novel methods to analyze the powers of the \emph{Coppersmith-Winograd tensor}, a basic construction introduced thirty years ago. In this paper we show how to generalize this approach to make progress on the complexity of rectangular matrix multiplication as well, by developing a framework to analyze powers of tensors in an asymmetric way. By applying this methodology to the fourth power of the Coppersmith-Winograd tensor, we succeed in improving the complexity of rectangular matrix multiplication. 

Let $\alpha$ denote the maximum value such that the product of an $n\times n^\alpha$ matrix by an $n^\alpha\times n$ matrix can be computed with $O(n^{2+\epsilon})$ arithmetic operations for any $\epsilon>0$. By analyzing the fourth power of the Coppersmith-Winograd tensor using our methods, we obtain the new lower bound $\alpha>0.31389$, which improves the previous lower bound $\alpha>0.30298$ obtained by Le Gall (FOCS'12) from the analysis of the second power of the Coppersmith-Winograd tensor. More generally, we give faster algorithms computing the product of an $n\times n^k$ matrix by an $n^k\times n$ matrix for any value $k\neq 1$. (In the case $k=1$, we recover the bounds recently obtained for square matrix multiplication). 

These improvements immediately lead to improvements in the complexity of a multitude of fundamental problems for which the bottleneck is rectangular matrix multiplication, such as computing the all-pair shortest paths in directed graphs with bounded weights.
\end{abstract}
\newpage
\section{Introduction}
\subsection{Background}
Matrix multiplication is one of the most important problems in mathematics and computer science. In 1969, Strassen discovered the first algorithm with subcubic complexity computing the product of two square matrices \cite{Strassen69}.  In modern notation, Strassen's result can be stated as an upper bound $\omega<2.81$ on the exponent of square matrix multiplication $\omega$, defined as the minimum value such that two $n\times n$ matrices can be multiplied using $O(n^{\omega+\epsilon})$ arithmetic operations for any constant $\epsilon<0$. Strassen's breakthrough initiated intense work on the complexity of matrix multiplication, which in a span of a few decades lead to several improvements, culminating in the celebrated $O(n^{2.376})$-time algorithm for square matrix multiplication by Coppersmith and Winograd \cite{Coppersmith+90}, i.e., the upper bound $\omega<2.376$ on the exponent of square matrix multiplication. This algorithm is obtained from a basic construction, which is nowadays often called the \emph{Coppersmith-Winograd tensor}. Coppersmith and Winograd showed that analyzing this tensor gives the upper bound $\omega<2.388$, and next showed that analyzing the second power of this tensor gives the improved upper bound $\omega<2.376$. 

A natural question, already mentioned in Coppersmith and Winograd's paper \cite{Coppersmith+90}, was whether higher powers of the Coppersmith-Winograd tensor can lead to further improvement to the complexity of matrix multiplication. Most efforts to investigate this direction quickly stopped after discovering that the third power does not seem to lead to any further improvement. More than twenty year later, however, Stothers \cite{Stothers10} (see also \cite{Davie+13}) and Vassilevska Williams \cite{WilliamsSTOC12} showed that the fourth power does give an improvement: the fourth power leads to the upper bound $\omega\le 2.373$. The technically challenging analysis of the fourth power was made possible by the introduction of powerful general recursive techniques to analyze powers of tensors. Extending these techniques, Vassilevska Williams \cite{WilliamsSTOC12} and then Le Gall~\cite{LeGallISSAC14} succeeded in analyzing higher powers up to the 32nd power, which gave additional small improvements and lead to the current best known upper bound on the exponent of square matrix multiplication $\omega<2.3728639$. Table \ref{table:chart} summarises all these results. Ambainis et al.~\cite{Ambainis+STOC15} finally showed that further improving this upper bound will be hard: they showed that analyzing higher powers of the Coppersmith-Winograd tensor  (e.g., powers 64, 128,...) using the same methodology cannot give any further significant improvement on $\omega$ (in particular it cannot lead to a proof of the popular conjecture $\omega=2$).

\begin{table}[b]
\renewcommand\arraystretch{1}
\begin{center}
\begin{tabular}{|c|l|l|}
\hline
$m$& Upper bound& Reference\bigstrut\\ 
\hline
1& $\omega<2.3871900$ & Coppersmith and Winograd~\cite{Coppersmith+90}\\
\hline
2&  $\omega<2.3754770$ & Coppersmith and Winograd~\cite{Coppersmith+90}\\
\hline
4& $\omega<2.372927$& Vassilevska Williams~\cite{WilliamsSTOC12}\\
\hline
8& $\omega<2.372873$ &  Vassilevska Williams~\cite{WilliamsSTOC12}\\
\hline
16& $\omega<2.3728640$ & Le Gall~\cite{LeGallISSAC14}\\
\hline
32& $\omega<2.3728639$ &  Le Gall~\cite{LeGallISSAC14}
\tabularnewline\hline
\end{tabular}
\end{center}
\vspace{-3mm}
\caption{Upper bounds on $\omega$ obtained by analyzing the $m$-th power of the Coppersmith and Winograd tensor.}\label{table:chart}
\end{table}

Besides square matrix multiplication, rectangular matrix multiplication plays a central role in many algorithms as well. In addition to natural applications to computational problems in linear algebra, typical examples of application include the construction of fast algorithms for the all-pairs shortest paths problem \cite{Alon+ESA07, Roditty+11,YusterSODA09,ZwickSTOC99,ZwickJACM02},  dynamic computation of the transitive closure \cite{Demetrescu+FOCS00,Sankowski+10}, detection of subgraphs \cite{Vassilevska+SODA15, Yuster+SODA04}, speed-up of sparse square matrix multiplication \cite{Amossen+09,Kaplan+06,Yuster+05} and algorithms for bounded-difference min-plus square matrix multiplication \cite{Bringmann+FOCS16}. Rectangular matrix multiplication has also been used in computational complexity \cite{Abboud+FOCS15,Patrascu+SODA10,WilliamsCCC11} and computational geometry~\cite{Kaplan+SODA07,Kaplan+06}.

The typical problem considered when studying rectangular matrix multiplication is computing the product of an $n\times \ceil{n^k}$ matrix by an $\ceil{n^k}\times n$ matrix, for some parameter $k\ge 0$.\footnote{Note that a basic result in algebraic complexity theory states that the algebraic  complexities of the following three problems are the same: computing the product of an $n\times \ceil{n^k}$ matrix by an $\ceil{n^k}\times n$ matrix, computing the product of an $\ceil{n^k}\times n$ matrix by an $n\times n$ matrix, and computing the product of an $n\times n$ matrix by an $n\times \ceil{n^k}$ matrix. In this paper for concreteness we discuss only the first type of products, but all our bounds naturally hold for the two other types as well.} 
In analogy to the square case, the exponent of rectangular matrix multiplication, denoted $\omega(k)$, is defined as the minimum value such that this product can be computed using $O(n^{\omega(k)+\epsilon})$ arithmetic operations for any constant $\epsilon>0$.
Also note that for $k=1$ (i.e., for square matrices), we have $\omega(1)=\omega$. 

Coppersmith~\cite{CoppersmithSICOMP82} showed in 1982 that $\omega(0.172)=2$. This surprising result means that the product of an $n\times \ceil{n^{0.172}}$ matrix by an $\ceil{n^{0.172}}\times n$ matrix can be computed in time almost linear in the size of the output (which contains $n^2$ entries). This discovery lead to the introduction of the following quantity $\alpha$:
\[
\alpha = \sup\{k\;|\:\omega(k)=2\}.
\]
Since proving that $\alpha=1$ is equivalent to proving that $\omega=2$, the quantity $\alpha$ is sometimes called the dual exponent of matrix multiplication. Coppersmith's result \cite{CoppersmithSICOMP82} then corresponds to the bound $\alpha\ge 0.172$. Coppersmith \cite{Coppersmith97} later showed that $\alpha>0.29462$ by analyzing the Coppersmith-Winograd tensor in the context of rectangular matrix multiplication. Fifteen years later, Le Gall \cite{LeGallFOCS12} showed that the second power of the Coppersmith-Winograd tensor can also be analyzed in the context of rectangular matrix multiplication, which lead to the improved lower bound $\alpha>0.30298$. This analysis was actually much more general and gave bounds on $\omega(k)$ that improved prior bounds \cite{Huang+98,Ke+08} for any $k\neq 1$. (For $k=1$, i.e., square matrix multiplication, this approach recovered the upper bound $\omega<2.376$ from~\cite{Coppersmith+90}). The results from \cite{LeGallFOCS12} are presented in Table \ref{table_results2}.

\begin{table}[t]
\begin{minipage}[t]{0.33\linewidth}\centering
\begin{tabular}{ |c | c |}
  \hline
  \multirow{2}{*}{$k$} & upper bound \\
  &on $\omega(k)$\\
  \hline                      
0.30298 &  2  \\
0.31 &   2.000063 \\
0.32 &   2.000371 \\
0.33 &   2.000939 \\
0.34 &   2.001771 \\
0.35 &   2.002870 \\
0.40 &   2.012175 \\
0.45 &   2.027102 \\
0.50 &   2.046681 \\
\hline  
\end{tabular}
\end{minipage}
\begin{minipage}[t]{0.33\linewidth}
\centering
\begin{tabular}{ |c | c |}
  \hline
  \multirow{2}{*}{$k$} & upper bound \\
  &on $\omega(k)$\\
  \hline
  0.5302&2.060396\\
0.55 &   2.070063 \\
  0.60 &   2.096571 \\
  0.65 &   2.125676 \\
  0.70 &  2.156959 \\
  0.75 &  2.190087 \\
0.80 & 2.224790 \\
  0.85 &  2.260830 \\
  0.90 &  2.298048 \\
0.95 &  2.336306 \\
1.00 &  2.375477 \\
  \hline  
\end{tabular}
\end{minipage}
\begin{minipage}[t]{0.33\linewidth}
\centering
\begin{tabular}{ |c | c | }
  \hline
  \multirow{2}{*}{$k$} & upper bound \\
  &on $\omega(k)$\\
  \hline    
1.10 &   2.456151 \\
1.20 &   2.539392 \\
1.30 &   2.624703 \\
1.40 &   2.711707 \\
1.50 &   2.800116 \\
1.75 &   3.025906 \\
2.00 &   3.256689 \\
2.50  & 3.727808 \\
3.00  & 4.207372 \\
4.00  &  5.180715 \\
5.00  & 6.166736 \\
  \hline  
\end{tabular}
\end{minipage}
\caption{Upper bounds from \cite{LeGallFOCS12} on the exponent of the multiplication of an 
$n\times n^k$ matrix by an $n^k\times n$ matrix, obtained by analyzing the second power of the Coppersmith-Winograd tensor.
\label{table_results2}}
\end{table}

\subsection{Our results}
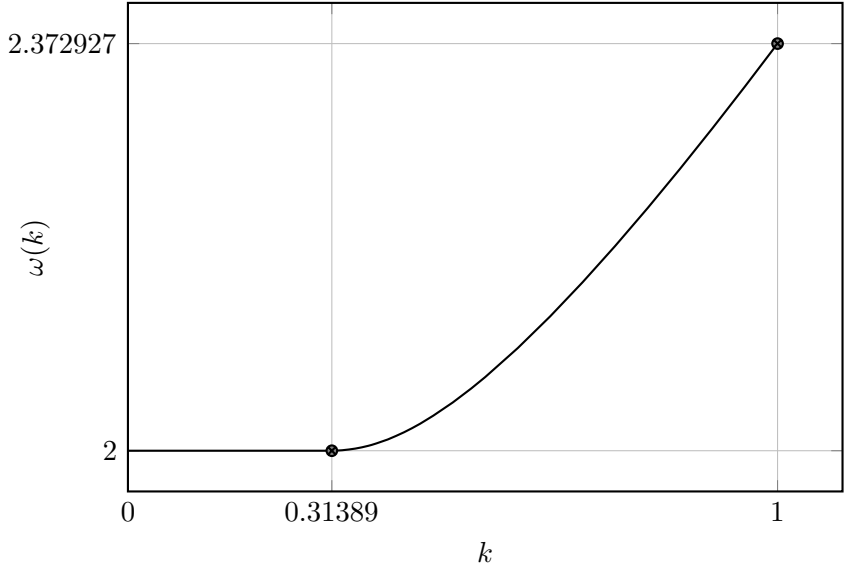
\begin{figure}[t]
\centering
\begin{tikzpicture}
\begin{axis}[
legend cell align=left,
width=9.5cm, 
height=6.5cm,
xmin=0, xmax=1.1,
thick,
 scale only axis,
xmajorgrids,
ymajorgrids,
 xtick={0,0.31389,1},
 xticklabels={0,0.31389,1},
  ytick={2,2.372927},
 yticklabels={2,2.372927},
xlabel={$k$},
ylabel={$\omega(k)$},
legend pos=south east]

\addplot [solid, every mark/.append style={solid, fill=gray}] coordinates {
(0,2)
(0.31477,2.000001)
(0.31508,2.000002)
(0.31532,2.000003)
(0.31552,2.000004)
(0.32,2.000064)
(0.33,2.000448)
(0.34,2.001118)
(0.35,2.001957)
(0.36,2.0031) 
(0.37,2.0045) 
(0.375,2.0053)
(0.3825,2.0067)
(0.4,2.010314)
(0.4125,2.0135)
(0.425,2.0169)
(0.4375,2.0207)
(0.45,2.024801)
(0.47,2.0321)
(0.5,2.044183)
(0.5286,2.057085)
(0.55,2.067488)
(0.6,2.093981)
(0.65,2.123097)
(0.7,2.154399)
(0.75,2.187543)
(0.8,2.222256)
(0.85,2.258317)
(0.9,2.295544) 
(0.95,2.333789)
(1,2.372927)
};

\addplot [only marks, every mark/.append style={solid, fill=gray}, mark=otimes*] coordinates {
(0.31389,2)
(1,2.372927)
};

\end{axis}
\end{tikzpicture}
\caption{\label{fig1}Our upper bounds on the exponent of the multiplication of an 
$n\times n^k$ matrix by an $n^k\times n$ matrix.}
\end{figure}

In view of the recent progress in square matrix multiplication algorithms obtained by analyzing higher powers of the Coppersmith-Winograd tensor, it is natural to ask whether the same approach can be applied to obtain further improvements on the complexity of rectangular matrix multiplication as well. We investigate this question in this paper, and present a framework to extend the analysis of higher powers to the case of rectangular matrix multiplication. We concretely focus on the analysis of the fourth power of the Coppersmith-Winograd tensor and show that this analysis leads to non-negligible improvements. The new upper bounds we obtain on the exponent of rectangular matrix multiplication $\omega(k)$ are given in Table \ref{table_results4} and the values for $k\le 1$ are plotted in Figure \ref{fig1}.\footnote{Note that the curve of Figure \ref{fig1} has the same shape as the curve for the second power given in \cite{LeGallFOCS12}.} We obtain in particular the new lower bound 
\[
\alpha \ge 0.31389
\]
on the dual exponent of matrix multiplication, as stated in the following theorem.

\begin{table}[t]
\begin{minipage}[b]{0.33\linewidth}\centering
\begin{tabular}{ |c | c |}
  \hline
  \multirow{2}{*}{$k$} & upper bound \\
  &on $\omega(k)$\\
  \hline                      
0.31389 & 2\\
0.32 & 2.000064\\
0.33 & 2.000448\\
0.34 & 2.001118\\
0.35 & 2.001957\\
0.40 & 2.010314\\
0.45 & 2.024801\\
0.50 & 2.044183\\
\hline  
\end{tabular}
\end{minipage}
\begin{minipage}[b]{0.33\linewidth}
\centering
\begin{tabular}{ |c | c |}
  \hline
  \multirow{2}{*}{$k$} & upper bound \\
  &on $\omega(k)$\\
  \hline
0.5286 & 2.057085\\
0.55 & 2.067488\\
0.60 & 2.093981\\
0.65 & 2.123097\\
0.70 & 2.154399\\
0.75 & 2.187543\\
0.80 & 2.222256\\
0.85 & 2.258317\\
0.90 & 2.295544\\
0.95 & 2.333789\\
1.00 & 2.372927\\
  \hline  
\end{tabular}
\end{minipage}
\begin{minipage}[b]{0.33\linewidth}
\centering
\begin{tabular}{ |c | c | }
  \hline
  \multirow{2}{*}{$k$} & upper bound \\
  &on $\omega(k)$\\
  \hline    
1.10 & 2.453481\\
1.20 & 2.536550\\
1.30 & 2.621644\\
1.40 & 2.708400\\
1.50 & 2.796537\\
1.75 & 3.021591\\
2.00 & 3.251640\\
2.50 & 3.721503\\
3.00 & 4.199712\\
4.00 & 5.171210\\
5.00 & 6.157233\\
  \hline  
\end{tabular}
\end{minipage}
\caption{Our upper bounds on the exponent of the multiplication of an 
$n\times n^k$ matrix by an $n^k\times n$ matrix, obtained by analyzing the fourth power of the Coppersmith-Winograd tensor.
\label{table_results4}}
\end{table}

\begin{theorem}\label{th:ma}
The product of an $n\times \ceil{n^{0.31389}}$ matrix by an $\ceil{n^{0.31389}}\times n$ matrix can by computed with $O(n^{2+\epsilon})$ arithmetic operations for any constant $\epsilon>0$.
\end{theorem}
This new bound improves the previous best known lower bound $\alpha>0.30298$ by Le Gall \cite{LeGallFOCS12}. For other values of $k$ as well, our new upper bounds on $\omega(k)$ are systematically better than those of \cite{LeGallFOCS12}, as can be seen by comparing Table \ref{table_results2} and Table \ref{table_results4}.  For instance we obtain ${\omega(3)\le 4.199712}$, which improves the previous upper bound ${\omega(3)\le 4.207372}$. Note that for $k=1$ (i.e., for square matrix multiplication), we obtain the same upper bound $\omega\le 2.372927$ on the exponent of square matrix multiplication as the bound obtained by analyzing the fourth power \cite{Davie+13,LeGallISSAC14,Stothers10,WilliamsSTOC12}. Indeed, for $k=1$ our analysis becomes essentially the same as the analysis for the square case in those prior works.

A surprising, or at least unexpected, aspect of the result of Theorem \ref{th:ma} is that the improvement from the second power to the fourth power (from $\alpha>0.30298$ to $\alpha>0.31389$) exceeds the improvement known from the first power to the second power (from $\alpha>0.29462$ to $\alpha>0.30298$). This is completely different from the improvements achieved on $\omega$ when analyzing successive powers of the Coppersmith-Winograd tensor, which are decreasing, as summarized in Table~\ref{table:chart}. Actually, all the numerical results we have obtained confirm that for any fixed value of $k$ the improvements on $\omega(k)$ decrease similarly to the square case when analyzing successive powers. For instance for $k=0.8$ and $k=2$ the first power gives $\omega(0.8)<2.2356$ and $\omega(2)<3.2699$; by examining Tables~\ref{table_results2} and \ref{table_results4} we observe that the improvement is larger from the first power to the second power. The situation happens to be different, however, for lower bounds on $\alpha$. Since the curves representing the upper bounds on $\omega(k)$ have horizontal asymptotes at the lower bound on $\alpha$ (see Figure \ref{fig1} of the present paper and Figure 1 in \cite{LeGallFOCS12}), even small improvements on $\omega(k)$ can lead to fairly significant improvements on $\alpha$, as our results show.

The most pressing question is now to investigate what will happen for even higher powers of the Coppersmith-Winograd tensor (e.g., power 8 or 16). We believe that this question is important since, besides its theoretical interest, further significant improvements for $\alpha$ may be obtained in this way. A  concrete approach would be to adapt to the rectangular case the numerically efficient methods based on convex optimization developed, in the setting of square matrix multiplication, to study high powers of the Coppersmith-Winograd tensor \cite{LeGallISSAC14}. In the other direction, it may be possible to show some limitations on the improvements achievable when studying higher powers, by generalizing the recent approach developed for the square case \cite{Ambainis+STOC15}.

\paragraph{Applications of our results.}
Our new bounds can be used to improve essentially all the known algorithms based on rectangular matrix multiplication algorithms (e.g., the algorithms in \cite{Alon+ESA07, Kaplan+SODA07,Kaplan+06, Roditty+11,YusterSODA09,Vassilevska+SODA15, Yuster+SODA04,ZwickSTOC99,ZwickJACM02}). Following \cite{LeGallFOCS12}, we discuss below one concrete example.

Zwick \cite{ZwickJACM02} has shown how to use rectangular matrix multiplication to compute the all-pairs shortest paths in weighted direct graphs where the weights are bounded integers. The time complexity obtained by Zwick for graphs with constant weights is $O(n^{2+\mu+\epsilon})$, for any constant $\epsilon>0$, where $\mu$ is the solution of the equation $\omega(\mu)=1+2\mu$. The results from \cite{LeGallFOCS12} (see Table \ref{table_results2}) show that $\omega(0.5302)<2.0604$, which gives the upper bound $\mu<0.5302$. The results of the present paper (see Table \ref{table_results4}) show that $\omega(0.5286)<2.0572$, which gives the upper bound $\mu<0.5286$.  

\subsection{Overview of our approach}
Before presenting an overview of the techniques used in this paper,
we give
an informal description of algebraic complexity theory 
(a more detailed presentation of these notions is given in Section~\ref{sec:prelim}).

\paragraph{Trilinear forms and the asymptotic sum inequality.}
The matrix multiplication of an $m\times n$ matrix by an $n\times p$ matrix can be represented by the following
trilinear form, denoted as $\braket{m,n,p}$:
\begin{equation*}\label{eq_mm}
\braket{m,n,p}=\sum_{r=1}^m\sum_{s=1}^n\sum_{t=1}^p x_{rs}y_{st}z_{rt},
\end{equation*}
where $x_{rs}$, $y_{st}$ and $z_{rt}$ are formal variables.
This form can be interpreted as follows: the $(r,t)$-th entry of the product of an $m\times n$ matrix $M$ by an $n\times p$ 
matrix $M'$ can be obtained by setting $x_{ij}=M_{ij}$ for all $(i,j)\in[m]\times[n]$ and $y_{ij}=M'_{ij}$ for all $(i,j)\in[n]\times[p]$, 
setting $z_{rt}=1$ and setting all the other $z$-variables to zero.
One can then think of the $z$-variables as formal variables used to record the entries of the matrix product.

More generally, a trilinear form $t$ is represented as 
$$
t=\sum_{u\in A}\sum_{v\in B}\sum_{w\in C} t_{uvw}x_{u}y_{v}z_{w}.
$$
where $A,B$ and $C$ are three sets, $x_{u}$, $y_{v}$ and $z_{w}$ are formal variables and
the $t_{uvw}$'s are coefficients in a field $\field$. The rank of the trilinear form $t$, denoted $R(t)$, represents the number of multiplications needed for the computation. The border rank of $t$, denoted $\underline{R}(t)$, is a generalization of the concept of rank, which is also related to the complexity of computing the tensor. 

A sum $\sum_{i}t_i$ of trilinear forms is a direct sum if the $t_i$'s do not
share variables. Informally, Sch\"onhage's asymptotic sum inequality \cite{Schonhage81} 
for rectangular matrix multiplication
states that, if the form $t$ can be converted 
into a direct sum of $c$ trilinear forms, 
each form being isomorphic
to $\braket{m,m,m^k}$, then 
$$
c\cdot m^{\omega(k)}\le \underline{R}(t).
$$
This implies that to obtain good upper bounds on the exponent of rectangular matrix multiplication, it is enough to find a tensor
$t$ of low border rank that can be converted into many independent (i.e., not sharing any variables) 
products of large enough rectangular matrices. 

\paragraph{Overview of the analysis of the second power.}
We now give a brief overview of the analysis of the second power of the Coppersmith-Winograd tensor given in \cite{LeGallFOCS12} to derive the upper bounds on $\omega(k)$ of Table \ref{table_results2}. 
The Coppersmith-Winograd tensor is a trilinear form $F_q$ introduced in \cite{Coppersmith+90}. Here $q$ is a parameter (concretely, $q$ is an integer between 2 and 10). Its second power $F_q\otimes F_q$ can actually be written as a sum of fifteen terms $T_{uvw}$:
$$
F_q\otimes F_q=\sum_{\begin{subarray}{c}0\le u,v,w\le 4\\ u+v+w=4\end{subarray}}T_{uvw}.
$$
In order to apply Sch\"onhage's asymptotic sum inequality, this sum must first be converted into a direct sum. This is done using a powerful general technique known as the laser method, first introduced by Strassen \cite{StrassenFOCS86} and then successively generalized and refined \cite{Coppersmith+90,Davie+13,Huang+98,Ke+08,LeGallFOCS12, LeGallISSAC14,Stothers10,WilliamsSTOC12}. The first step is to take the $N$-th tensor product of the basic construction, where $N$ is a large integer, and then zero variables so that the remaining terms do not share variables. Since we want each remaining term to be isomorphic to a rectangular matrix product in order to obtain an upper bound on $\omega(k)$ via Sch\"onhage's asymptotic sum inequality, the choice of zeroed variables has to be done carefully. The laser method allows us, for any choice of the fifteen parameters $a_{uvw}\in[0,1]$ satisfying specific constraints, to convert the $N$-th tensor product of the basic construction into a direct sum of many terms (the number of these terms depending on the values of the $a_{uvw}$'s), each isomorphic to
\begin{equation}\label{eq_T}
\bigotimes_{\begin{subarray}{c}0\le u,v,w\le 4\\ u+v+w=4\end{subarray}}T_{uvw}^{\otimes a_{uvw}N}.
\end{equation}

The next step is to analyze each term (\ref{eq_T}) and show that it corresponds to a direct sum of matrix products of the form $\braket{m,m,m^k}$ (the number of terms in the direct sum and the value of $m$ will depend on the values of the $a_{uvw}$'s and $q$). Some of the $T_{uvw}$'s (more precisely, all the $T_{uvw}$'s except $T_{112}$, $T_{121}$ and $T_{211}$) can be analyzed in a straightforward way, since they correspond to matrix products. The main technical contribution of the approach from \cite{LeGallFOCS12} was to show that each of the remaining three terms can be converted into a large number of objects called ``$\Cc$-tensors'' in Strassen's terminology \cite{Strassen87}. This conversion is done again via the laser method, which introduces additional parameters. Finally,  Ref.~\cite{LeGallFOCS12} explained how to convert these $\Cc$-tensors into a direct sum of matrix multiplication tensors. Combining the analysis of these fifteen terms shows that~(\ref{eq_T}) corresponds to a direct sum of matrix products of the form $\braket{m,m,m^k}$, as wanted. Sch\"onhage's asymptotic sum inequality then gives the upper bound on $\omega(k)$ presented in Table \ref{table_results2} by numerically optimizing the choice of the parameters (the choice of $q$, the $a_{uvw}$'s and the additional parameters arising in the second extraction).

\paragraph{Overview of our analysis of the fourth power.}
The fourth power of the Coppersmith-Winograd tensor  can be written as a sum of 45 terms $T_{uvw}$:
\[
F_q^{\otimes 4}=\sum_{\begin{subarray}{c}0\le u,v,w\le 8\\ u+v+w=8\end{subarray}}T_{uvw}.
\]
For conciseness, this tensor will be denoted $F$ through the paper. Similarly to the analysis of the second power, the laser method allows us, for any choice of parameters $a_{uvw}\in[0,1]$ satisfying specific constraints, to convert the $N$-th tensor product of the basic construction into a direct sum of many terms, each isomorphic to
\begin{equation}\label{eq_T2}
\bigotimes_{\begin{subarray}{c}0\le u,v,w\le 8\\ u+v+w=8\end{subarray}}T_{uvw}^{\otimes a_{uvw}N}.
\end{equation}
We call this process the \emph{first extraction}, which is explained in detail in Section \ref{sec:extra1}. Note that while this extraction is more complicated than for the second power since the number of variables is larger and deriving the constraints that the parameters should satisfy is more complex, conceptually the analysis is fairly standard.

The main technical contribution of this work is a methodology to analyze each term (\ref{eq_T2}). A natural strategy would be to mimic the analysis done in \cite{LeGallFOCS12} for the second power and analyze each component $T_{uvw}$ individually. While this leads to some improvement over the second power when $k$ is close to 1 (in particular, this leads to the same upper bound $\omega<2.372927$ as in prior works analyzing the fourth power in the context of square matrix multiplication \cite{Stothers10,WilliamsSTOC12}), this strategy does not give any improvement for smaller values of $k$ (in particular, no improved lower bound on the dual exponent of matrix multiplication $\alpha$). Our strategy, instead, is to analyze all the terms $T_{uvw}$ together via the laser method. As in the term-by-term analysis done for the second power in \cite{LeGallFOCS12}, this introduces a set of new parameters for each term and a set of constraints that these parameters should satisfy. A difference is that now some of the constraints are global: they can involve the parameters of all the 45 terms. We call this process the \emph{second extraction}, which is explained in detail in Section \ref{sec:extra2}. Note that this methodology appears to be more powerful than the term-by-term conversion to $\Cc$-tensors done in \cite{LeGallFOCS12}: First, as already mentioned, the latter approach does not seem to lead to any improvement on $\alpha$ for the fourth power. Second, our new methodology, when applied to the analysis of the second power in replacement of the conversion into $\Cc$-tensors done in \cite{LeGallFOCS12}, already leads to upper bounds on $\omega(k)$ slightly better than those found in \cite{LeGallFOCS12} for some values of~$k$ (more precisely, we observed such improvements for values in the range $k\in[0.37,0.46]$). 

The second extraction outlined in the previous paragraph actually does not completely analyze~(\ref{eq_T2}): it simply decomposes each term $T_{uvw}$ into a direct sum of products of the fifteen terms arising in the analysis of the second power. To complete the analysis, we recursively apply the same strategy as for the second extraction and analyse the contribution of all these fifteen terms together, again using the laser method (which introduce two additional parameters). We call this process the \emph{third extraction}, which is explained in detail in Section \ref{sec:extra3}. 

Finally, combining our three extractions, we conclude that the tensor $F^{\otimes N}$ can be converted into a direct sum of $c$ trilinear forms, 
each form being isomorphic to $\braket{m,m,m^k}$, for some values $c$ and $m$ depending on all the parameters introduced. Applying Sch\"onhage's asymptotic sum inequality then gives an inequality involving $\omega(k)$ and all these parameters (the formal statement is Theorem~\ref{thm:main} in Section \ref{sec:full}). Optimizing numerically the choice of parameters, under the constraints derived on those parameters, gives the upper bounds of Table \ref{table_results4} and the lower bound on $\alpha$ of Theorem \ref{th:ma}. 
\section{Preliminaries}\label{sec:prelim}

We present various known results and tools related to matrix multiplication. Two good references for an extensive treatment of this topic are \cite{Burgisser+97} and \cite{Blaser13}.

\subsection{General notations and definitions}
\begin{sloppypar}
In this paper we will use, for any positive integer $n$, the notation $\Ni{1}{n}$ to represent the set $\{1,\ldots,n\}$.
Given a finite set $X$ we denote $\Pp(X)$ the set of all function $a\colon X\to ]0,1[\,\cap\,\Q$ such that ${\sum_{x\in X}a(x)=1}$. Note that the functions in $\Pp(X)$ are simply probability distributions on $X$ where the probabilities are rational and strictly between 0 and 1.\\
\end{sloppypar}

We define the notion of \textit{type}. A type can be seen as a frequency vector.
\begin{definition}\label{def:type}
Given $N\in\N$, a finite set $U$, two mappings 
$u:~\Ni{1}{N} \longrightarrow U$ and $t\in\Pp(U)$,
we say that $u$ is of type $t$ if 
$|u^{-1}(x)|=t(x)N$ holds for all $x \in U$.
\end{definition}

\subsection{Tensors, matrix multiplication and the asymptotic sum inequality}\label{subsec:trilin}
Let $\F$ be a field, let $u$,$v$,$w$ be three positive integers, and let $U=\F^u$, $V=\F^v$ and $W=\F^w$.
A tensor $t$ of format $(u,v,w)$ is an element of $U\otimes V\otimes W=\F^{u\times v\times w}$,
where $\otimes$ denotes the tensor product.
Fix a base $(x_i)$ of $U$, a base $(y_j)$ of $V$ and a base $(z_k)$ of $W$.
We write $x_iy_jz_k$ as a short cut for $x_i\otimes y_j\otimes z_k$. The family $(x_iy_jz_k)$ is a base of $U\otimes V\otimes W$, and thus $t$ can be written in this base as
$t=\sum\limits_{ijk}t_{ijk}x_iy_jz_k$, where the $t_{ijk}$ are coefficients in $\F$. For a tensor written under this form, we call the $(x_i)$ the $x$ variables, and similarly we call the $(y_j)$ and the $(z_k)$ the $y$ variables and the $z$ variables.

Let $t\in U\otimes V\otimes W$ and $t'\in U'\otimes V'\otimes W'$ be two tensors.
The direct sum $t\oplus t'$, is a tensor in 
$(U\oplus U')\otimes (V\oplus V')\otimes (W\oplus W')$.
The tensor product $t\otimes t'$ is a tensor in 
$(U\otimes U')\otimes (V\otimes V')\otimes (W\otimes W')$. 
For any positive $c$, we will denote the tensor $t\oplus\cdots\oplus t$ (with $c$ occurrences of $t$)
by $c\cdot t$ and 
the tensor $t\otimes\cdots \otimes t$ (with $c$ occurrences of $t$)
by $t^{\otimes c}$.\\

The tensor representing the multiplication of an $m \times n$ matrix by an $n \times p$ matrix over the field~$\F$, denoted $\braket{m,n,p}$, is the tensor of format $(mn,np,mp)$ defined as
$$\braket{m,n,p} = \sum\limits_{ijk}t_{ijk}x_iy_jz_k$$
where $i$ spans $\Ni{1}{m} \times \Ni{1}{n}$, $j$ spans $\Ni{1}{n} \times \Ni{1}{p}$, $k$ spans $\Ni{1}{m} \times \Ni{1}{p}$ and $$
t_{ijk}=
\left\{\!
\begin{array}{ll}
1&\textrm{if } i=(r,s), j=(s,t) \textrm{ and } k=(r,t) \textrm{ for some integers } 
(r,s,t)\in \Ni{1}{m} \times \Ni{1}{n}\times \Ni{1}{p}\\
0&\textrm{otherwise}.
\end{array}
\right.
$$

We also consider the tensor of format $(n,n,n)$ which represents $n$ independent scalar products. It is denoted by $\braket{n}$ and is defined as $\braket{n}=\sum\limits_{l=1}^n x_l y_l z_l$.

The following definitions allow us to relate the properties of different tensors. Let $\lambda$ denote an indeterminate, and $\F[\lambda]$ the space of polynomials in $\lambda$ with coefficients in $\F$.

\begin{definition} 
Let $t \in \F^{u\times v\times w}$ and $t' \in \F[\lambda]^{u'\times v'\times w'}$ be two tensors. We say that $t'$ is a restriction of $t$, and denote $t' \le t$, if there are linear maps $\alpha: \F^u \longrightarrow \F[\lambda]^{u'}$, $\beta: \F^v \longrightarrow \F[\lambda]^{v'}$, $\gamma: \F^w \longrightarrow \F[\lambda]^{w'}$ such that $(\alpha\otimes\beta\otimes\gamma)(t) = t'$ where $(\alpha\otimes\beta\otimes\gamma)$ is the linear map $\F^u\otimes\F^v\otimes\F^w \longrightarrow \F[\lambda]^{u'}\otimes\F[\lambda]^{v'}\otimes\F[\lambda]^{w'}$ obtained by taking the tensor product of $\alpha$, $\beta$, $\gamma$.
\end{definition}

The intuition behind this notion is that the \textit{restriction} of a tensor is easier to compute than the original tensor, in the sense that an algorithm computing a tensor $t$ can be converted into an algorithm computing a tensor $t' \le t$ with the same complexity.

\begin{definition}
Let $t \in \F[\lambda]^{u\times v\times w}$ and $t' \in \F^{u\times v\times w}$ be two tensors.
We say that $t'$ is an approximation of $t$, if there exists a tensor 
$t'' \in \F[\lambda]^{u\times v\times w}$ and some $s\in\N$ such that 
$t=\lambda^s t' + \lambda^{s+1} t''$. We may also write : $t = \lambda^s t' + \calO(\lambda^{s+1})$.
\end{definition}

This is analogous to the notion of approximate computation.

\begin{definition}
Let $t \in \F^{u\times v\times w}$ and $t' \in \F^{u'\times v'\times w'}$ be two tensors.
We say that $t'$ is a degeneration of $t$, and denote $t' \degen t$, if $t'$ is an approximation of a restriction of $t$.
\end{definition}

Note that by definition, $t' \le t \Longrightarrow t' \degen t$. The notion of degeneration can be seen as an approximate conversion. It has the following property.

\begin{proposition}[Proposition 15.25 in \cite{Burgisser+97}]\label{prop_fact}
Let $t_1,t_1',t_2$ and $t_2'$ be four tensors.
Suppose that $t'_1\unlhd t_1$ and $t'_2\unlhd t_2$.
Then $t'_1\oplus t'_2\unlhd t_1\oplus t_2$ and $t'_1\otimes t'_2\unlhd t_1\otimes t_2$.
\end{proposition}

\begin{definition}
Let $t$ be a tensor. The border rank of $t$ is $\underline{R}(t)=\min\{{r\in\N\:|\: t\degen\braket{r}\}}$.
\end{definition}

The notion of border rank enables us to formally define the exponent of rectangular matrix multiplication, as follows. For any $k\ge 0$, 
\[
\omega(k)=\inf\{\beta\:|\:\underline{R}(\braket{n,\sfloor{n^k},n})=O(n^\beta)\}.
\]
The exponent of square matrix multiplication is $\omega=\omega(1)$.

Similarly to almost all recent works on matrix multiplications, our main tool for proving lower bounds on $\omega(k)$ will be Sch\"onhage's asymptotic sum inequality \cite{Schonhage81} (see \cite{Huang+98,Ke+08,LeGallFOCS12,Lotti+83} for the version of the inequality given below). 

\begin{theorem}[Sch\"onhage's asymptotic sum inequality] \label{thm:Schonhage}
Let $k$, $m$ and $c$ be three positive integers.
Let $t$ be a tensor such that $c\cdot \braket{m,m,m^k}\unlhd t$.
Then 
$$c\cdot  m^{\omega(k)}\le \underline{R}(t).$$
\end{theorem}

\subsection{The fourth power of Coppersmith-Winograd tensor}\label{subsec:CWtensor}
\begin{sloppypar}For any positive integer $q$, the Coppersmith-Winograd tensor \cite{Coppersmith+90} is the tensor of format ${(q+2,q+2,q+2)}$ defined as\end{sloppypar}
$$F_q=\sum_{i=1}^q (x_0y_iz_i+x_iy_0z_i+x_iy_iz_0)+x_0y_0z_{q+1}+x_0y_{q+1}z_{0}+x_{q+1}y_0z_0.$$
Coppersmith and Winograd showed that $\underline{R}(F_q)\le q+2$.

They also considered the square of this tensor. For any $t\in\N$, define the set 
$$S_t=\{(i,j,k)\in\N^3 \mid i+j+k = t\}.$$ Define the tensors $x_{ii'} y_{jj'} z_{kk'} = x_i y_j z_k \otimes x_{i'} y_{j'} z_{k'}$.
By regrouping terms, we can write
$$(F_q)^{\otimes 2} = \sum_{(i,j,k) \in S_4} T_{ijk}$$
where
\begin{eqnarray*}
T_{004}&=&x_{0,0}y_{0,0}z_{q+1,q+1}\\
T_{013}&=&\sum_{i=1}^q x_{0,0}y_{i,0}z_{i,q+1}+\sum_{k=1}^q x_{0,0}y_{0,k}z_{q+1,k}\\
T_{022}&=&x_{0,0}y_{q+1,0}z_{0,q+1}+x_{0,0}y_{0,q+1}z_{q+1,0}+\sum_{i,k=1}^q x_{0,0}y_{i,k}z_{i,k}\\
T_{112}&=&\sum_{i=1}^q x_{i,0}y_{i,0}z_{0,q+1}+\sum_{k=1}^q x_{0,k}y_{0,k}z_{q+1,0}+\sum_{i,k=1}^q x_{i,0}y_{0,k}z_{i,k}+\sum_{i,k=1}^q x_{0,k}y_{i,0}z_{i,k}
\end{eqnarray*}
and the other eleven terms are obtained by permuting the indexes of the $x$ variables, the $y$ variables and $z$ variables in the above expressions (e.g., $T_{040}=x_{0,0}y_{q+1,q+1}z_{0,0}$ and $T_{400}=x_{q+1,q+1}y_{0,0}z_{0,0}$). Note that $\underline{R}(F_q^{\otimes 2}) \le (q+2)^2$ from the submultiplicativity of the border rank.

Let us now consider the fourth power of the Coppersmith-Winograd tensor $F_q$ (already studied, in the context of square matrix multiplication, in Refs.~\cite{Davie+13,Stothers10,LeGallISSAC14,WilliamsSTOC12}).
For any $(i,j,k)\in S_8$ define the set 
$$S_{ijk} = \{((u,v,w),(u',v',w'))\in S_4\times S_4 \mid u+u'=i, v+v'=j, w+w'=k\}$$
and the tensor 
$$T_{ijk} = \sum_{((u,v,w),(u',v',w')) \in S_{ijk}} T_{uvw}\otimes T_{u'v'w'}.$$
By regrouping terms, the fourth power of $F_q$, which hereafter we simply denote $F$ (the value of $q$ will be implicit until the very end of the paper), can be written as
\begin{align*}
F &= (F_q)^{\otimes 4}\\
&= (F_q)^{\otimes 2} \otimes (F_q)^{\otimes 2}\\
&= \left(\sum_{(u,v,w)\in S_4} T_{uvw}\right) \otimes \left(\sum_{(u',v',w')\in S_4} T_{u'v'w'}\right)\\
&= \sum_{((u,v,w),(u',v',w'))\in S_4^2} T_{uvw}\otimes T_{u'v'w'}\\
&= \sum_{(i,j,k) \in S_8} \:\:\:\:
\sum_{((u,v,w),(u',v',w'))\in S_{ijk}} T_{uvw}\otimes T_{u'v'w'}\\
&= \sum_{(i,j,k) \in S_8} T_{ijk}. 
\end{align*}
Note that $\underline{R}(F) \le (q+2)^4$, again from the submultiplicativity of the border rank.

When later working with the terms $T_{ijk}$, we will sometimes consider the equivalent decomposition
$$T_{ijk} = \sum_{(u,v,w) \in \overline{S}_{ijk}} V_{ijk}[uvw]$$
where
$$\overline{S}_{ijk} = \{(u,v,w)\in S_4, \exists~(u',v',w')\in S_4, u+u'=i, v+v'=j, w+w'=k\}$$
and 
$$\forall~(u,v,w)\in \overline{S}_{ijk}, V_{ijk}[uvw]=T_{uvw}\otimes T_{i-u,j-v,k-w}.$$

For any integer $N\in\N$ and triple $(I,J,K)$ with ${I=(I(1),\dots,I(N))}$, $J=(J(1),\dots,J(N))$, $K=(K(1),\dots,K(N))$ such that $(I(1),J(1),K(1)),\dots,(I(N),J(N),K(N))\in S_8$, define
$$T_{IJK}=T_{I(1) J(1) K(1)}\otimes\dots\otimes T_{I(N) J(N) K(N)}.$$
Notice that we define $T_{IJK}$ only for triple $(I,J,K)$ from the set
$$\{(I,J,K), \forall~l\in\Ni{1}{N}, (I(l),J(l),K(l))\in S_8\} \cong S_8^N.$$
We can then write: $$F^{\otimes N} = \sum_{(I,J,K)\in S_8^N} T_{IJK}.$$
Finally, for any triple $(a,b,c)\in S_8$ and any triple
$(I,J,K) = (I(l),J(l),K(l))_{l\in\Ni{1}{N}} \in \overline{S}_{abc}^N$, we define
$$V_{abc}[IJK]=\bigotimes\limits_{l=1}^{N}V_{abc}[I(l) J(l) K(l)].$$

\subsection{Extraction from a tensor}\label{subsec:thextraction}

In this subsection we explain our main tool to realize an \textit{extraction} from a sum of tensors. An extraction consists in assigning some variables to zero in a tensor (thus eliminating all their contributions to the sum). If a tensor $T'$ is extracted from $T$, then $T' \le T$ trivially holds. Our primary goal is to guarantee that the resulting tensor $T'$ is a direct sum of isomorphic tensors, so that the asymptotic sum inequality can be used.

All recent progresses on square or rectangular matrix multiplication have been obtained by performing extractions based on the so-called laser method \cite{Coppersmith+90,Davie+13,Huang+98,Ke+08,LeGallFOCS12, LeGallISSAC14,Stothers10,StrassenFOCS86,WilliamsSTOC12}. Le Gall \cite{LeGallFOCS12} introduced the following convenient framework to interpret such reductions in the rectangular case. In this framework, a sum of tensors corresponds to a graph whose vertices are the tensors in the sum. There is an edge between two vertices in the graph if and only if the two corresponding terms in the sum of tensors share a variable. Let $G$ denote this graph, and $U$ denote its set of vertices.  Zeroing a term in the sum corresponds to removing one vertex from the graph. As mentioned above, however, terms can be zeroed only by zeroing the variables it contains. This means that such a zeroing operation may actually remove more than one vertex from the graph. Extracting a direct sum from the original tensor is then equivalent to removing vertices from the graph by such zeroing operations and reaching an edgeless graph. When using this methodology, we will like to additionally guarantee that the vertices remaining in the final graph are from a specified  subset $U^\ast\subseteq U$. Concretely, the set $U^*$ will be the set of vertices of terms matching a certain type, which will ensure that all the tensors remaining after the extraction are of this type. In our extractions we will use the following theorem from \cite{LeGallFOCS12}, which is tailored for this goal and was already used for the analysis of the second power of the Coppersmith-Winograd tensor. 

\begin{theorem}[Theorem 4.2 in \cite{LeGallFOCS12}]\label{thm:extraction}
Let $\tau$ be a fixed positive integer.
Let $N$ be a large integer and  
define the set 
$$\Lambda= \left\{(I,J,K)\in [\tau]^N\times [\tau]^N\times [\tau]^N \:|\: I_\ell+J_\ell+K_\ell=\tau \textrm{ for all } \ell\in\{1,\ldots,N\}\right\}.$$
Define the three coordinate functions $f_1,f_2,f_3\colon [\tau]^N\times [\tau]^N\times [\tau]^N\to [\tau]^N$ as follows.
\begin{eqnarray*}
f_1((I,J,K))&=&I\\
f_2((I,J,K))&=&J\\ 
f_3((I,J,K))&=&K
\end{eqnarray*}
Let $U$ be a subset of $\Lambda$ such that there exist integers $\Nn_1,\Nn_2$ and $\Nn_3$ for which
the following property holds: for any $I\in [\tau]^N$, 
\begin{eqnarray*}
|\{u\in U \:|\: f_1(u)=I\}|&\in&\{0,\Nn_1\}\\
|\{u\in U \:|\: f_2(u)=I\}|&\in&\{0,\Nn_2\}\\
|\{u\in U \:|\: f_3(u)=I\}|&\in&\{0,\Nn_3\}.
\end{eqnarray*}
Let $\calT_1=|f_1(U)|$, $\calT_2=|f_2(U)|$ and $\calT_3=|f_3(U)|$.
Let $G$ be the (simple and undirected) graph with vertex set $U$ in which two distinct vertices~$u$ and~$v$ are connected if and only if there exists one index $i\in\{1,2,3\}$ such that $f_i(u)=f_i(v)$.

Assume there exists a set $U^\ast\subseteq U$ such that 
\begin{itemize}
\item
$|f_i(U^\ast)|=\calT_i$ for each $i\in\{1,2,3\}$;
\item
there exist integers $\Nn^\ast_1,\Nn^\ast_2$ and $\Nn^\ast_3$ such that
$$|\{u\in U \:|\: f_i(u)=I\}|=\Nn_i \Leftrightarrow |\{u\in U^\ast \:|\: f_i(u)=I\}|=\Nn_i^\ast$$
for each $I\in [\tau]^N$ and each $i\in\{1,2,3\}$.
\end{itemize}
Define a removal operation as removing all the vertices $u$ (if any) such that $f_i(u)=I$, for a fixed sequence $I\in [\tau]^N$ and a fixed position $i\in\{1,2,3\}$
Then, for any constant $\epsilon>0$, the graph $G$ can be converted, with only removal operations, into an edgeless graph with 
$$\Omega\left(\frac{\calT_1\Nn^\ast_1}{(\Nn_1+\Nn_2+\Nn_3)^{1+\epsilon}}\right)$$ vertices, all of them being in $U^\ast$.
\end{theorem}
\section{First extraction}\label{sec:extra1}
In this section we describe our first extraction.

Let us consider any function $a\in\Pp(S_8)$. For any $(i,j,k)\in S_8$ we will often write $a(ijk)$ instead of $a(i,j,k)$.
Given $a$, we define the following three mappings:
$$A:~\Ni{0}{8} \longrightarrow~]0,1[~~~~i\longmapsto A(i)=\sum_{j,k\in\N\mid (i,j,k)\in S_8} a(ijk),$$ 
$$B:~\Ni{0}{8} \longrightarrow~]0,1[~~~~j\longmapsto B(j)=\sum_{i,k\in\N\mid (i,j,k)\in S_8} a(ijk),$$
$$C:~\Ni{0}{8} \longrightarrow~]0,1[~~~~k\longmapsto C(k)=\sum_{i,j\in\N\mid (i,j,k)\in S_8} a(ijk).$$
$A$,$B$,$C$ are the projections of $a$ on each of the three coordinates. We thus have
$$\sum_{i\in\Ni{0}{8}} A(i)=\sum_{j\in\Ni{0}{8}} B(j)=\sum_{k\in\Ni{0}{8}} C(k)=1.$$

\begin{sloppypar}
We are going to realize a first extraction from the tensor $F^{\otimes N}$, where $N$ is such that 
${\forall~(i,j,k)\in S_8, a(ijk)N\in\N}$ (such an $N$ always exists since the values of $a$ are rational). The goal is to zero variables so that, from the sum 
\[
F^{\otimes N}=\sum\limits_{(I,J,K)\in S_8^N} T_{IJK},
\] 
we are left only with tensors $T_{IJK}$ where $(I,J,K)$ is of type $a$, i.e, tensors isomorphic to 
\[
\bigotimes\limits_{(i,j,k) \in S_8} T_{ijk}^{a(ijk)N}.
\]
\end{sloppypar}
We now explain how to achieve this goal.

\paragraph{The extraction.}
First notice that for any ${(I,J,K),(I',J',K')\in S_8^N}$ such that $I\neq I'$, the $x$ variables in $T_{IJK}$ and in $T_{I'J'K'}$ are disjoint. We set to zero all the $x$ variables except the ones which appear in a $T_{IJK}$ where $I$ is of type $A$, that is to say we set to zero all variables which appear in a $T_{IJK}$ with $I$ not of type $A$. We are thus left with only the tensors $T_{IJK}$ with $I$ of type $A$.

We apply the same process for the $y$ and $z$ variables so that only remain the tensors $T_{IJK}$ with $I$ of type $A$, $J$ of type $B$ and $K$ of type $C$. 
Let $[a]$ denote the set of mappings $\overline{a}:~S_8\longrightarrow~]0,1[$ which have the same projections as $a$ and satisfy 
$\forall~(i,j,k)\in S_8,~\overline{a}(ijk)N\in\N$. The tensors which remain are exactly the tensors $T_{IJK}$ with $(I,J,K)$ of type $\overline{a}\in[a]$.

The number of sequences $I\in\Ni{0}{8}^N$ of type $A$ is 
$$\calT_X={N\choose{(A(i) N)_{i\in\Ni{0}{8}}}}$$
as choosing a sequence $I$ of type $A$ is equivalent to choose the location of the $A(i)$ elements $i$ for $i\in\Ni{0}{8}$. Using the Stirling formula, we get, with the $A(i)$ fixed and $N \longrightarrow \infty$,
$$\calT_X=\Theta\left(\frac{1}{N^{4}}\left(\frac{1}{\prod\limits_{i=0}^8 A(i)^{A(i)}}\right)^{N}\right).$$
Similarly, we define the number of sequences $J\in\Ni{0}{8}^N$ of type $B$ as $\calT_Y$ and the number of sequences $K\in\Ni{0}{8}^N$ of type $C$ as $\calT_Z$.

For any fixed sequence $I$ of type $A$, the number of remaining forms $T_{IJK}$ of type $a$ is
$$\calN^{*}_X=\prod\limits_{i=0}^8 
{A(i) N\choose{(a(ijk)N)_{j,k\in\N\mid(i,j,k)\in S_8}}},$$
while the total number of remaining forms $T_{IJK}$ is
$$\calN_X=\sum_{\overline{a}\in[a]}\prod\limits_{i=0}^8 
{A(i) N\choose{(\overline{a}(ijk)N)_{j,k\in\N\mid(i,j,k)\in S_8}}}.$$
Define, the function $g$ which associates to any mapping
$x:~S_8 \longrightarrow~]0,1[~~(i,j,k)\longmapsto x(ijk)$ the value
$g(x)=\left(\prod\limits_{(i,j,k)\in S_8}x(ijk)^{x(ijk)}\right)^{-1}$.
Using Stirling's formula, and the fact that $|S_8|=\sum\limits_{l=1}^9 l=45$, we get that
$$\calN_X^{*}=\Theta\left(\dfrac{\left[g(a)\prod\limits_{i=0}^8 A(i)^{A(i)}\right]^N}{N^{18}}\right),\:\:\:
\calN_X=\Theta\left(\sum_{\overline{a}\in[a]}
\dfrac{\left[g(\overline{a})\prod\limits_{i=0}^8 A(i)^{A(i)}\right]^N}{N^{18}}\right).$$
Similarly, for any sequence $J$ of type $B$, the number of remaining forms $T_{IJK}$ of type $a$ and the total number of remaining forms $T_{IJK}$ are
$$\calN_Y^{*}=\Theta\left(\dfrac{\left[g(a)\prod\limits_{j=0}^8 B(j)^{B(j)}\right]^N}{N^{18}}\right),\:\:\:
\calN_Y=\Theta\left(\sum_{\overline{a}\in[a]}
\dfrac{\left[g(\overline{a})\prod\limits_{j=0}^8 B(j)^{B(j)}\right]^N}{N^{18}}\right),$$
and for any sequence $K$ of type $C$, the number of remaining forms $T_{IJK}$ of type $a$ and the total number of remaining forms $T_{IJK}$ are
$$\calN_Z^{*}=\Theta\left(\dfrac{\left[g(a)\prod\limits_{k=0}^8 C(k)^{C(k)}\right]^N}{N^{18}}\right),\:\:\:
\calN_Z=\Theta\left(\sum_{\overline{a}\in[a]}
\dfrac{\left[g(\overline{a})\prod\limits_{k=0}^8 C(k)^{C(k)}\right]^N}{N^{18}}\right).$$

Using the framework presented in Subsection \ref{subsec:thextraction}, we get by Theorem \ref{thm:extraction} that for any $\epsilon>0$ we can further extract from the remaining $T_{IJK}$ a direct sum of 
$$\Omega\left(\frac{\calT_X\calN^\ast_X}{(\calN_X+\calN_Y+\calN_Z)^{1+\epsilon}}\right)$$
tensors $T_{IJK}$, all of which are of type $a$. We want the number of tensors is this direct sum to be high. For this to happen, we now formulate some conditions on $a$.\\

\paragraph{Conditions on $a$.}
We first want to find $a\in \Pp(S_8)$ such that $g(a)=\max\limits_{\overline{a}\in[a]}g(\overline{a})$. Let us see $g$ as a function from the set
$]0,1[^{45}$ to the set of positive real numbers. We want to find the maximum of $g$ on the domain 
$[a] \subset ]0,1[^{45}$. For this, we will find the maximum of $g$ on the domain 
$\widetilde{[a]}$, the set of mappings $S_8\longrightarrow~]0,1[$ which have the same projections as $a$.
Note that $\widetilde{[a]} \supseteq [a]$, and that $\widetilde{[a]}$ is a convex subset of $]0,1[^{45}$.
Consider the function $\ln g:~\widetilde{[a]}\longrightarrow \R$. For any $\overline{a}\in\widetilde{[a]}$,
 $\ln g (\overline{a}) = \sum\limits_{(i,j,k)\in S_8} -\overline{a}(ijk)\ln \overline{a}(ijk)$. 
Since $\ln g $ is a concave function on a convex domain, any critical point of $\ln g$ is a global maximum of $\ln g$, and thus of $g$. We express the conditions satisfied by the critical points below.

We first observe that $\ln g$ can actually be written as a (concave) function of only $21$ variables, namely
$\overline{a}(215)$, $\overline{a}(224)$, $\overline{a}(233)$, $\overline{a}(242)$, $\overline{a}(251)$, $\overline{a}(260)$, $\overline{a}(314)$, $\overline{a}(323)$, $\overline{a}(332)$, $\overline{a}(341)$, $\overline{a}(350)$, $\overline{a}(413)$, $\overline{a}(422)$, $\overline{a}(431)$, $\overline{a}(440)$, $\overline{a}(512)$, $\overline{a}(521)$, $\overline{a}(530)$, $\overline{a}(611)$, $\overline{a}(620)$, $\overline{a}(710)$. This is because $\ln g$ is defined on $\widetilde{[a]}$, and the elements of $\widetilde{[a]}$ have, by definition, the same projections as $a$, and thus for any $\overline{a}\in\widetilde{[a]}$ the $\overline{a}(ijk)$ satisfy the following system of linear equations:\\
\begin{align*}
\forall~i\in\Ni{0}{8}, ~A(i)&=\sum_{j,k\in\N\mid (i,j,k)\in S_8} \overline{a}(ijk),\\
\forall~j\in\Ni{0}{8}, ~B(j)&=\sum_{i,k\in\N\mid (i,j,k)\in S_8} \overline{a}(ijk),\\
\forall~k\in\Ni{0}{8}, ~C(k)&=\sum_{i,j\in\N\mid (i,j,k)\in S_8} \overline{a}(ijk).\\
\end{align*}
Resolving of the (homogeneous) linear system\footnote{A Maple file deriving the symbolic solution of this system is available at \cite{fileurl}.} reduces the number of variables to 21, as claimed. 

From now we will assume that $a$ satisfies the symmetry condition 
\begin{equation}
\tag{C1}
\forall~(i,j,k)\in S_8, a(ijk)=a(ikj).
\label{eqn:symm1}
\end{equation}
Computing each one of the $21$ partial differential equations $\dfrac{\partial \ln g}{\partial\overline{a}(ijk)} = 0$ leads, after simplification and by condition $\eqref{eqn:symm1}$, to a system of $10$ non linear equations:
\begin{align*}
	0 &= - \ln(\overline{a}(017)) + \ln(\overline{a}(026)) + \ln(\overline{a}(107)) - \ln(\overline{a}(125)) - \ln(\overline{a}(206)) + \ln(\overline{a}(215)),\\
	0 &= - \ln(\overline{a}(017)) + \ln(\overline{a}(026)) + \ln(\overline{a}(107)) - \ln(\overline{a}(116)) - \ln(\overline{a}(602)) + \ln(\overline{a}(611)),\\
	0 &= - \ln(\overline{a}(017)) + \ln(\overline{a}(035)) + \ln(\overline{a}(107)) - \ln(\overline{a}(134)) - \ln(\overline{a}(305)) + \ln(\overline{a}(314)),\\
	0 &= - \ln(\overline{a}(017)) + \ln(\overline{a}(044)) + \ln(\overline{a}(107)) - \ln(\overline{a}(134)) - \ln(\overline{a}(404)) + \ln(\overline{a}(413)),\\
	0 &= - \ln(\overline{a}(017)) + \ln(\overline{a}(035)) + \ln(\overline{a}(107)) - \ln(\overline{a}(125)) - \ln(\overline{a}(503)) + \ln(\overline{a}(512)),\\
	0 &= - \ln(\overline{a}(017)) + \ln(\overline{a}(035)) + \ln(\overline{a}(107)) + \ln(\overline{a}(116)) - \ln(\overline{a}(125)) - \ln(\overline{a}(134)) -\\
	&~~~~~~~\ln(\overline{a}(206)) + \ln(\overline{a}(224)),\tag{C2}\label{eqn:xi}\\
	0 &= - \ln(\overline{a}(017)) + \ln(\overline{a}(044)) + \ln(\overline{a}(107)) + \ln(\overline{a}(116)) - \ln(\overline{a}(134)) - \ln(\overline{a}(134)) -\\
	&~~~~~~~\ln(\overline{a}(206)) + \ln(\overline{a}(233)),\\
	0 &= - \ln(\overline{a}(017)) - \ln(\overline{a}(026)) + \ln(\overline{a}(035)) + \ln(\overline{a}(044)) + \ln(\overline{a}(107)) + \ln(\overline{a}(116)) -\\
	&~~~~~~~\ln(\overline{a}(134)) - \ln(\overline{a}(134)) - \ln(\overline{a}(305)) + \ln(\overline{a}(323)),\\
	0 &= - \ln(\overline{a}(017)) - \ln(\overline{a}(026)) + \ln(\overline{a}(044)) + \ln(\overline{a}(035)) + \ln(\overline{a}(107)) + \ln(\overline{a}(116)) -\\
	&~~~~~~~\ln(\overline{a}(134)) - \ln(\overline{a}(134)) - \ln(\overline{a}(305)) + \ln(\overline{a}(323)),\\
	0 &= - \ln(\overline{a}(017)) - \ln(\overline{a}(026)) + \ln(\overline{a}(044)) + \ln(\overline{a}(035)) + \ln(\overline{a}(107)) + \ln(\overline{a}(116)) - \\
	&~~~~~~~\ln(\overline{a}(134)) - \ln(\overline{a}(125)) - \ln(\overline{a}(404)) + \ln(\overline{a}(422)).
\end{align*}
\begin{sloppypar}Note that, as the value of any $\overline{a}\in[a]$ is fixed from  the values of only $21$ variables, and as 
${\forall~(i,j,k)\in S_8, a(ijk)N\in\Ni{1}{N}}$, we have $|[a]|\le N^{21}$. For any $a$ satisfying these $10$ equations, we have 
$g(a)=\max\limits_{\overline{a}\in[a]} g(\overline{a})$ and thus, 
$\calN_X=\calO(N^{21}\calN^*_X)$, $\calN_Y=\calO(N^{21}\calN^*_Y)$, $\calN_Z=\calO(N^{21}\calN^*_Z)$.\\\end{sloppypar}

\paragraph{Final statement.}
Let us also impose the condition 
\[
\tag{C3}
\prod\limits_{i=0}^8 A(i)^{A(i)} \ge \prod\limits_{j=0}^8 B(j)^{B(j)}. 
\label{eqn:C3}
\] 
\begin{sloppypar}By the symmetry condition $\eqref{eqn:symm1}$, this implies that
$\prod\limits_{i=0}^8 A(i)^{A(i)} \ge \prod\limits_{k=0}^8 C(k)^{C(k)}$. We get ${\calN^*_Y=\calN^*_Z=\calO(\calN^*_X)}$, and thus we obtain
${(\calN_X+\calN_Y+\calN_Z)=\calO(N^{21}\calN^*_X)}$
and\end{sloppypar}
$$\frac{\calT_X\calN^\ast_X}{(\calN_X+\calN_Y+\calN_Z)^{1+\epsilon}}=
\Omega\left(\frac{\calT_X}{(N^{21(1+\epsilon)}(\calN^*_X)^{\epsilon}}\right).$$
As by definition $\calN^*_X \le |S_8|^N=45^N$, this is equal to
$$r_1=\Omega\left(\dfrac{1}{N^{25+21\epsilon}45^{N\epsilon}}
\left[\dfrac{1}{\prod\limits_{i=0}^8 A(i)^{A(i)}}\right]^N\right).$$
We obtain the following final result.
\begin{theorem}\label{th:extra1}
Let $q$ be any positive integer. Let $a$ be any function from $\Pp(S_8)$ satisfying the constraints $\eqref{eqn:symm1}$, $\eqref{eqn:xi}$ and $\eqref{eqn:C3}$.
Then for any $\epsilon > 0$, the trilinear form $F^{\otimes N}$ admits a restriction which is a direct sum of $r_1$
trilinear forms, each of which is isomorphic to
$\bigotimes\limits_{(i,j,k)\in S_8} T_{ijk}^{\otimes a(ijk)N}$.
\end{theorem}
\section{Second extraction}\label{sec:extra2}

As we will see later in details in Section \ref{sec:full}, the tensors $T_{ijk}$ for $(i,j,k)\in S_8$ with one or more of their indices $i,j,k$ equal to $0$ are actually matrix products tensors. No further work is required for them. In contrast, the tensors $T_{ijk}$ for $(i,j,k)\in\overline{S}_8$ where $\overline{S}_8 = \{(i,j,k) \in S_8 \mid i>0, j>0, k>0\}$ do not correspond to matrix products.

We are now going to realize an extraction on all the tensors $T_{ijk}, (i,j,k)\in \overline{S}_8$. 
We first study the properties of the $T_{ijk}, (i,j,k)\in \overline{S}_8$. In Subsection \ref{subsec:T233}, we consider the particular case of the tensors $T_{233}$, $T_{323}$ and $T_{332}$. In Subsection \ref{subsec:remaining_tensors}, we consider the remaining tensors, i.e., the tensors $T_{ijk}$ for 
${(i,j,k) \in S'_8 = \overline{S}_8 \setminus \{(2,3,3), (3,2,3), (3,3,2)\}}$, which are actually easier to analyze. Then, in Subsection \ref{subsec:joint}, we explain the limitations of independent extractions and introduce our method to realize a joint extraction.

\subsection{The tensors {\boldmath$T_{233}$}, {\boldmath$T_{323}$}, {\boldmath$T_{332}$}}\label{subsec:T233}

The extractions from the tensors $T_{233}$, $T_{323}$, $T_{332}$ can be realized similarly to the extraction from the tensor $F$ that we realized in Section \ref{sec:extra1}. As the situation is similar for the three tensors, we only detail the extraction from the tensor $T_{233}$.

\begin{sloppypar}We start from the decomposition
\[
T_{233} = \sum\limits_{(u,v,w) \in \overline{S}_{233}} V_{233}[uvw].
\]
 By definition of the $(T_{ijk})_{(i,j,k)\in S_4}$, for any $(i,j,k),(i',j',k')\in \overline{S}_{233}$ such that $i\neq i'$, the $x$ variables in $T_{ijk}$ and in $T_{i'j'k'}$ are disjoints, and thus the $x$ variables in $V_{233}[ijk]$ and in $V_{233}[i'j'k']$ are disjoint. We can thus realize an extraction from $T_{233}^{a(233)N}$ just as in Section \ref{sec:extra1}. We consider a mapping 
$a_{233}\in\Pp(\overline{S}_{233})$, with projections $A_{233}$, $B_{233}$, $C_{233}$.
As the $(a_{233}(ijk)a(233))$ are rational numbers, and as $N$ will later go to infinity, we can assume that 
$\forall~(i,j,k)\in \overline{S}_{233},~a_{233}(ijk)a(233)N\in\N$. 
We impose the symmetry condition 
$\forall~(i,j,k)\in \overline{S}_{233}, a_{233}(ijk)=a_{233}(ikj)$, and thus $B_{233}=C_{233}$. We also impose the symmetry condition $$\forall~(i,j,k)\in \overline{S}_{233}, a_{233}(ijk) = a_{233}(2-i,3-j,3-k),$$ 
as $V_{233}[ijk] = V_{233}[2-i,3-j,3-k]$.
By realizing an extraction successively on the $x$ variables, the $y$ variables and the $z$ variables, we are left with the tensors $V_{233}[IJK]$ where ${(I,J,K)\in \overline{S}_{233}^{a(233)N}}$ is of type $\overline{a}_{233}\in[a_{233}]$. Here $[a_{233}]$ denotes the set of mappings 
$\overline{a}_{233}:~\overline{S}_{233} \longrightarrow~]0,1[$ which have the same projections as $a_{233}$ and satisfy
$\forall~(i,j,k)\in \overline{S}_{233}, \overline{a}_{233}(ijk)a(233)N\in\N$.\\\end{sloppypar}

The number of sequences $I\in\Ni{0}{2}^{a(233)N}$ of type $A_{233}$ is 
$$\calT_{233,X}={a(233)N\choose{(A_{233}(i) a(233)N)_{i\in\Ni{0}{2}}}}=
\Theta\left(\frac{1}{N}\left(\frac{1}{\prod\limits_{i=0}^2 A_{233}(i)^{A_{233}(i)}}\right)^{a(233)N}\right).$$

Define the function $g_{233}$ which associates to any mapping
$x:~\overline{S}_{233} \longrightarrow~]0,1[~~(i,j,k)\longmapsto x(ijk)$ the value
\[
g_{233}(x)=\left(\prod\limits_{(i,j,k)\in \overline{S}_{233}}x(ijk)^{x(ijk)}\right)^{-1}.
\]
For any fixed sequence $I$ of type $A_{233}$, the number of remaining forms $V_{233}[IJK]$ with $(I,J,K)$ of type $a_{233}$ is
$$\calN_{233,X}^*=\prod\limits_{i=0}^2
{A_{233}(i) a(233)N\choose{(a_{233}(ijk)a(233)N)_{j,k\in\N\mid(i,j,k)\in \overline{S}_{233}}}}=
\Theta\left(\dfrac{\left[g_{233}(a_{233})\prod\limits_{i=0}^2 A_{233}(i)^{A_{233}(i)}\right]^{a(233)N}}{(a(233)N)^{7/2}}\right),$$
while the total number of remaining forms $V_{233}[IJK]$ is
\begin{align*}
\calN_{233,X}&=\sum_{\overline{a}_{233}\in[a_{233}]}\prod\limits_{i=0}^2 
{A_{233}(i) N\choose{(\overline{a}_{233}(ijk)a(233)N)_{j,k\in\N\mid(i,j,k)\in \overline{S}_{233}}}}\\
&=\Theta\left(\sum_{\overline{a}_{233}\in[a_{233}]}
\dfrac{\left[g_{233}(\overline{a}_{233})\prod\limits_{i=0}^2 A_{233}(i)^{A_{233}(i)}\right]^{a(233)N}}{(a(233)N)^{7/2}}\right).
\end{align*}

Similarly, for any fixed sequence $J$ of type $B_{233}$, the number of remaining forms $V_{233}[IJK]$ with $(I,J,K)$ of type $a_{233}$ is
$$\calN_{233,Y}^*=
\Theta\left(\dfrac{\left[g_{233}(a_{233})\prod\limits_{j=0}^3 B_{233}(j)^{B_{233}(j)}\right]^{a(233)N}}{(a(233)N)^3}\right),$$
the total number of remaining forms $V_{233}[IJK]$ is
$$\calN_{233,Y}=\Theta\left(\sum_{\overline{a}_{233}\in[a_{233}]}
\dfrac{\left[g_{233}(\overline{a}_{233})\prod\limits_{j=0}^3 B_{233}(j)^{B_{233}(j)}\right]^{a(233)N}}{(a(233)N)^3}\right),$$
and for any fixed sequence $K$ of type $C_{233}$, the number of remaining forms $V_{233}[IJK]$ with $(I,J,K)$ of type $a_{233}$ is
$\calN_{233,Z}=\calN_{233,Y}$ and the total number of remaining forms $V_{233}[IJK]$ is
${\calN_{233,Z}^*=\calN_{233,Y}^*}$.

\begin{sloppypar}By studying the function $g_{233}$ in a similar way as we studied the function $g$ in Section \ref{sec:extra1}, we get that $g(a_{233})=\max\limits_{\overline{a}_{233}\in[a_{233}]} g(\overline{a}_{233})$ for $a_{233}$ satisfying the constraint
$$2\ln(a_{233}(211)) + \ln(a_{233}(130)) - \ln(a_{233}(202)) - \ln(a_{233}(220)) - \ln(a_{233}(112)) = 0.$$
Calculations show that ${\overline{a}_{233}\in[a_{233}]}$ can be written as a function of $\overline{a}_{233}(130)$ and $\overline{a}_{233}(103)$ only, and as 
${\overline{a}_{233}(130),\overline{a}_{233}(103)\in\Ni{1}{a(233)N}}$, we have that ${|[a_{233}]|\le(a(233)N)^2}$ and thus we obtain
${\calN_{233,X}=\calO((a(233)N)^2\calN_{233,X}^*)}$,
${\calN_{233,Y}=\calO((a(233)N)^2\calN_{233,Y}^*)}$ and
${\calN_{233,Z}=\calO((a(233)N)^2\calN_{233,Z}^*)}$.\\\end{sloppypar}

The tensors $T_{323}$ and $T_{332}$ are analysed similarly. Imposing the constraints  
\begin{align*}
2\ln(a_{323}(121)) + \ln(a_{323}(310)) - \ln(a_{323}(022)) - \ln(a_{323}(220)) - \ln(a_{323}(112)) &= 0\\
2\ln(a_{332}(112)) + \ln(a_{332}(031)) - \ln(a_{332}(202)) - \ln(a_{332}(022)) - \ln(a_{332}(211)) &= 0
\end{align*}
implies that $\calN_{323,X}=\calO(N^2\calN_{323,X}^*)$ and $\calN_{332,X}=\calO(N^2\calN_{332,X}^*)$.\\

To summarize, when analyzing $T_{233}$, $T_{323}$ and $T_{332}$ we need to impose the following three constraints (in addition to other constraints discussed later):
\begin{align*}
2\ln(a_{233}(211)) + \ln(a_{233}(130)) - \ln(a_{233}(202)) - \ln(a_{233}(220)) - \ln(a_{233}(112)) &= 0\\
2\ln(a_{323}(121)) + \ln(a_{323}(310)) - \ln(a_{323}(022)) - \ln(a_{323}(220)) - \ln(a_{323}(112)) &= 0
\tag{D2}\label{eqn:xi2}\\
2\ln(a_{332}(112)) + \ln(a_{332}(031)) - \ln(a_{332}(202)) - \ln(a_{332}(022)) - \ln(a_{332}(211)) &= 0.
\end{align*}

\subsection{The tensors of {\boldmath$S'_8$}}\label{subsec:remaining_tensors}

\begin{sloppypar}We adopt the same notations as in the previous subsection. We consider as before mappings 
${a_{ijk} \in \Pp(\overline{S}_{ijk})}$ for $(i,j,k) \in S'_8$, with projections $A_{ijk}$, $B_{ijk}$ and $C_{ijk}$, and assume as before ${\forall~(u,v,w)\in \overline{S}_{ijk},~a_{ijk}(uvw)a(ijk)N\in\N}$.
For $(i,j,k)\in S'_8$, the number of sequences $I\in\Ni{0}{i}^{a(ijk)N}$ of type $A_{ijk}$ is\end{sloppypar}
$$\calT_{ijk,X}={a(ijk)N\choose{(A_{ijk}(u) a(ijk)N)_{u\in\Ni{0}{i}}}}=
\Theta\left(\frac{1}{N^{\frac{i}{2}}}\left(\frac{1}{\prod\limits_{u=0}^i A_{ijk}(u)^{A_{ijk}(u)}}\right)^{a(ijk)N}\right).$$
\begin{sloppypar}The extractions on the elements of $S'_8$ are simpler in the sense that we have
${\forall~(i,j,k)\in S'_8, [a_{ijk}] = \{a_{ijk}\}}$, i.e., fixing the projections $A_{ijk}$, $B_{ijk}$ and $C_{ijk}$ fixes all the values of the mapping $a_{ijk}$. We thus have
\[
\calN_{ijk,X}=\calN^*_{ijk,X}, \:\:\:
\calN_{ijk,Y}=\calN^*_{ijk,Y}, \:\:\:
\calN_{ijk,Z}=\calN^*_{ijk,Z} ,
\]
with
\begin{align*}
\calN_{ijk,X}&= 
\prod\limits_{x=0}^i
{A_{ijk}(x) a(ijk)N\choose{(a_{ijk}(xyz)a(ijk)N)_{y,z\in\N\mid(x,y,z)\in \overline{S}_{ijk}}}}=
\Theta\left(\dfrac{\left[g_{ijk}(a_{ijk})\prod\limits_{x=0}^i A_{ijk}(x)^{A_{ijk}(x)}\right]^
{a(ijk)N}}{(a(ijk)N)^{\dfrac{|\overline{S}_{ijk}|-(i+1)}{2}}}\right),
\end{align*}
\begin{align*}
\calN_{ijk,Y}&= 
\prod\limits_{y=0}^j
{A_{ijk}(y) a(ijk)N\choose{(a_{ijk}(xyz)a(ijk)N)_{x,z\in\N\mid(x,y,z)\in \overline{S}_{ijk}}}}=
\Theta\left(\dfrac{\left[g_{ijk}(a_{ijk})\prod\limits_{y=0}^j B_{ijk}(y)^{B_{ijk}(y)}\right]^
{a(ijk)N}}{(a(ijk)N)^{\dfrac{|\overline{S}_{ijk}|-(j+1)}{2}}}\right),
\end{align*}
\begin{align*}
\calN_{ijk,Z}&= 
\prod\limits_{z=0}^k
{A_{ijk}(z) a(ijk)N\choose{(a_{ijk}(xyz)a(ijk)N)_{x,y\in\N\mid(x,y,z)\in \overline{S}_{ijk}}}}=
\Theta\left(\dfrac{\left[g_{ijk}(a_{ijk})\prod\limits_{z=0}^k C_{ijk}(z)^{C_{ijk}(z)}\right]^
{a(ijk)N}}{(a(ijk)N)^{\dfrac{|\overline{S}_{ijk}|-(k+1)}{2}}}\right).
\end{align*}
\end{sloppypar}

\subsection{The joint extraction}\label{subsec:joint}

If we were to realize the extraction on each of the $T_{ijk}, (i,j,k)\in \overline{S}_8$ independently, we would have to impose either the constraints 
$$\forall~(i,j,k)\in \overline{S}_8, 
\prod\limits_{x=0}^i A_{ijk}(x)^{A_{ijk}(x)} \ge \prod\limits_{y=0}^j B_{ijk}(y)^{B_{ijk}(y)}~\textrm{ and }~
\prod\limits_{x=0}^i A_{ijk}(x)^{A_{ijk}(x)} \ge \prod\limits_{z=0}^k C_{ijk}(z)^{C_{ijk}(z)}$$ 
or, as in \cite{LeGallFOCS12}, introduce the notion of $\Cc$-tensor to perform the analysis. It does not seem, however, that any of these two approaches is helpful for analyzing the fourth power of the Coppersmith-Winograd tensor in the rectangular setting (in particular, the above constraints are too strong and do not lead to any improved lower bound on $\alpha$). Instead, we are going to realize a global extraction directly on the tensor 
\begin{equation}\label{eq:tensorS8}
\bigotimes\limits_{(i,j,k)\in \overline{S}_8} T_{ijk}^{\otimes a(ijk)N}.
\end{equation}
Note that this tensor can be decomposed as follows:

\begin{align*}
\bigotimes\limits_{(i,j,k)\in \overline{S}_8} T_{ijk}^{\otimes a(ijk)N} &= 
\bigotimes\limits_{(i,j,k)\in \overline{S}_8} \left(\sum_{(u,v,w) \in \overline{S}_{ijk}} V_{ijk}[uvw]\right)^{\otimes a(ijk)N}\\
&=\bigotimes\limits_{(i,j,k)\in \overline{S}_8} \left(\sum_{(I,J,K) \in \overline{S}_{ijk}^{a(ijk)N}} 
V_{ijk}[IJK]\right)
\end{align*}
where for $(I,J,K) \in \overline{S}_{ijk}^{a(ijk)N}$, 
$V_{ijk}[IJK] = \bigotimes\limits_{l\in\Ni{1}{a(ijk)N}} V_{ijk}[I(l)J(l)K(l)].$
$$
\bigotimes\limits_{(i,j,k)\in \overline{S}_8} T_{ijk}^{\otimes a(ijk)N}
=\sum\limits_{(I,J,K) \in \prod\limits_{(i,j,k) \in \overline{S}_8} \overline{S}_{ijk}^{a(ijk)N}}
\left(\bigotimes\limits_{(i,j,k)\in \overline{S}_8} V_{ijk}[I_{ijk} J_{ijk} K_{ijk}]\right)$$
where we see $(I,J,K) \in \prod\limits_{(i,j,k) \in \overline{S}_8} \overline{S}_{ijk}^{a(ijk)N}$ 
as a family $(I_{ijk},J_{ijk},K_{ijk})_{(i,j,k)}$ indexed by $(i,j,k) \in \overline{S}_8$ with $(I_{ijk},J_{ijk},K_{ijk}) \in \overline{S}_{ijk}^{a(ijk)N}$.

Let $(i,j,k)\in \overline{S}_8$. By definition of the $(T_{uvw})_{(u,v,w)\in S_4}$, for any $(u,v,w),(u',v',w')\in \overline{S}_{ijk}$ such that $u\neq u'$, the $x$ variables in $T_{uvw}$ and in $T_{u'v'w'}$ are disjoint, and thus the $x$ variables in $V_{ijk}[uvw]$ and in $V_{ijk}[u'v'w']$ are disjoint. This implies that for any 
$(I_{ijk},J_{ijk},K_{ijk}),(I'_{ijk},J'_{ijk},K'_{ijk}) \in \overline{S}_{ijk}^{a(ijk)N}$ with $I_{ijk}\neq I'_{ijk}$, the $x$ variables in $V_{ijk}[I_{ijk} J_{ijk} K_{ijk}]$ and in $V_{ijk}[I'_{ijk} J'_{ijk} K'_{ijk}]$ are disjoint.

The $x$ variables of the tensor (\ref{eq:tensorS8}) are indexed by a sequence of indices. This sequence can be divided in a partition of subsequences, each subsequence being associated to a 
$V_{ijk}[I_{ijk},J_{ijk},K_{ijk}]$ for a $(i,j,k)\in \overline{S}_8$. 
Hence, for any $(I,J,K),(I',J',K')\in \prod_{(i,j,k) \in \overline{S}_8} \overline{S}_{ijk}^{a(ijk)N}$ with $I \neq I'$, the $x$ variables in 
$V[IJK]=\bigotimes_{(i,j,k)\in \overline{S}_8} V_{ijk}[I_{ijk} J_{ijk} K_{ijk}]$ and
$V[I'J'K']=\bigotimes_{(i,j,k)\in \overline{S}_8} V_{ijk}[I'_{ijk} J'_{ijk} K'_{ijk}]$ are disjoint.

We rewrite the tensor (\ref{eq:tensorS8}) as 
$$\bigotimes\limits_{(i,j,k)\in \overline{S}_8} T_{ijk}^{\otimes a(ijk)N} = 
\sum\limits_{(I,J,K) \in \prod\limits_{(i,j,k) \in \overline{S}_8} \overline{S}_{ijk}^{a(ijk)N}} V[IJK].$$

\begin{sloppypar}Define $\widetilde{a}=(a_{ijk})_{(i,j,k) \in \overline{S}_8}$ and the three projections
$\widetilde{A}=(A_{ijk})_{(i,j,k) \in \overline{S}_8}$,
$\widetilde{B}=(B_{ijk})_{(i,j,k) \in \overline{S}_8}$,
${\widetilde{C}=(C_{ijk})_{(i,j,k) \in \overline{S}_8}}$.
We extend the definition of type (Definition \ref{def:type}) to a product of types: we say, 
for ${(I,J,K) \in \prod_{(i,j,k) \in \overline{S}_8} \overline{S}_{ijk}^{a(ijk)N}}$,
that $I$ is of \textit{multi-type} $\widetilde{A}$ if for every $(i,j,k) \in \overline{S}_8$, $I_{ijk}$ is of type $A_{ijk}$, and that $(I,J,K)$ is of multi-type $\widetilde{a}$ if for every $(i,j,k) \in \overline{S}_8$ $(I_{ijk},J_{ijk},K_{ijk})$ is of type $a_{ijk}$.\end{sloppypar}

We set to $0$ all the $x$ variables except those that appear in a $V[IJK]$ where $I$ is of multi-type $\widetilde{A}$. We apply the same procedure for the $y$ variables with the multi-type $\widetilde{B}$ and for the $z$ variables with the multi-type $\widetilde{C}$. We are thus left with only the tensors $V[IJK]$ with $I$ of multi-type $\widetilde{A}$, $J$ of multi-type $\widetilde{B}$, $K$ of multi-type $\widetilde{C}$.

By definition of a multi-type, the number of $I$ of multi-type $\tilde{A}$ is
$$\widetilde{\calT_X} = \prod\limits_{(i,j,k)\in \overline{S}_8} \calT_{ijk,X}.$$
For any fixed $I$ of multi-type $\tilde{A}$, the number of remaining tensors $V[IJK]$ with $(I,J,K)$ of multi-type $\tilde{a}$ is
$$\widetilde{\calN}^*_X = \prod\limits_{(i,j,k)\in \overline{S}_8} \calN^*_{ijk,X}.$$
For any fixed $I$ of multi-type $\tilde{A}$, the number of remaining tensors $V[IJK]$ is
$$\widetilde{\calN}_X = \prod\limits_{(i,j,k)\in \overline{S}_8} \calN_{ijk,X}.$$
For any fixed $J$ of multi-type $\tilde{B}$, the number of remaining tensors $V[IJK]$ is
$$\widetilde{\calN}_Y = \prod\limits_{(i,j,k)\in \overline{S}_8} \calN_{ijk,Y}.$$
For any fixed $K$ of multi-type $\tilde{C}$, the number of remaining tensors $V[IJK]$ is
$$\widetilde{\calN}_Z = \prod\limits_{(i,j,k)\in \overline{S}_8} \calN_{ijk,Z}.$$

\begin{sloppypar}We now show that we are under the conditions of Theorem \ref{thm:extraction}. We set
$$\Lambda' = \left\{(I,J,K) \in \prod\limits_{(i,j,k) \in \overline{S}_8} S_4^{a(ijk)N}\right\}.$$
Note that as $\forall~(i,j,k)\in\overline{S}_8,~\overline{S}_{ijk}^{a(ijk)N} \subseteq S_4^{a(ijk)N}$, for any remaining tensor $V[IJK]$, we have that $(I,J,K) \in \Lambda'$.
Let $\gamma = \sum_{(i,j,k)\in \overline{S}_8} a(ijk)$. By defining an arbitrary ordering
${\theta: (i,j,k,l) \longmapsto \Ni{1}{\gamma N}}$ for ${(i,j,k) \in \overline{S}_8}$ and $l \in \Ni{1}{a(ijk)N}$, we have
$\Lambda' \cong \Lambda$ where\end{sloppypar}
$$\Lambda= 
\Big\{(I,J,K) \in \Ni{0}{4}^{\gamma N} \times \Ni{0}{4}^{\gamma N} \times \Ni{0}{4}^{\gamma N} \mid \forall~d \in \Ni{1}{\gamma N}, 
I[\theta^{-1}(d)] + J[\theta^{-1}(d)] + K[\theta^{-1}(d)] = 4 \Big\}$$ with the notation
$I[(i,j,k,l)] = I_{ijk}(l)$, $J[(i,j,k,l)] = J_{ijk}(l)$, $K[(i,j,k,l)] = K_{ijk}(l)$.

We set $U$ to be set of $(I,J,K)$ of the remaining $V[IJK]$. We set 
$\calN_1=\widetilde{\calN_X}$, $\calN_2=\widetilde{\calN_Y}$, $\calN_3=\widetilde{\calN_Z}$.
We set $U^*$ to be the set of $(I,J,K)$ of the remaining $V[IJK]$ with $(I,J,K)$ of multi-type $\widetilde{a}$. We set
$\calN^*_1=\widetilde{\calN^*_X}$, $\calN^*_2=\widetilde{\calN^*_Y}$, $\calN^*_3=\widetilde{\calN^*_Z}$.
We set $\calT_1=\widetilde{\calT_X}$, $\calT_2=\widetilde{\calT_Y}$, $\calT_3=\widetilde{\calT_Z}$, and all the conditions of Theorem \ref{thm:extraction} are satisfied. Fix $\epsilon > 0$. 
We obtain, from $\bigotimes\limits_{(i,j,k)\in \overline{S}_8} T_{ijk}^{\otimes a(ijk)N}$, a direct sum of
$$\Omega\left(\dfrac{\widetilde{T_X}\widetilde{\calN_X^*}}
{(\widetilde{\calN_X}+\widetilde{\calN_Y}+\widetilde{\calN_Z})^{1+\epsilon}}\right)$$
trilinear forms, each of which being a $V[IJK]$ with $(I,J,K)$ of multi-type $\widetilde{a}$, i.e., isomorphic to 
$$\bigotimes\limits_{(u,v,w)\in \overline{S}_8}\left(\bigotimes\limits_{(i,j,k) \in \overline{S}_{uvw}}
V_{uvw}(ijk)^{\otimes a_{uvw}(ijk)a(uvw)N}\right).$$

We now impose the constraint 
\begin{equation}
\tag{D3}
\prod\limits_{(i,j,k)\in \overline{S}_8} \left(\prod\limits_{x=0}^i A_{ijk}(x)^{A_{ijk}(x)}\right)^{a(ijk)} \ge 
\prod\limits_{(i,j,k)\in \overline{S}_8} \left(\prod\limits_{y=0}^j B_{ijk}(y)^{B_{ijk}(y)}\right)^{a(ijk)}.
\label{eqn:D3}
\end{equation}
This implies that 
$$\prod\limits_{(i,j,k)\in \overline{S}_8} \calN_{ijk,Y} = \calO\left(\prod\limits_{(i,j,k)\in \overline{S}_8} \calN_{ijk,X}\right)$$
i.e., $\widetilde{\calN_Y} = \calO (\widetilde{\calN_X})$. 

We impose the symmetry condition
\begin{align*}
\tag{D1}
\forall~(i,j,k)\in \overline{S}_8, \forall~(u,v,w) \in \overline{S}_{ijk}, 
a_{ijk}(uvw)=a_{ijk}(i-u,j-v,k-w)=a_{ikj}(u,w,v).
\label{eqn:symm2}
\end{align*}
The symmetry conditions $\eqref{eqn:symm1}$ and $\eqref{eqn:symm2}$ give $\widetilde{\calN_Y}=\widetilde{\calN_Z}$ and $\widetilde{\calN_Z} = \calO (\widetilde{\calN_X})$.
The number of terms in the direct sum that we obtain is
$$\Omega\left(\dfrac{\widetilde{\calT_X}\widetilde{\calN_X^*}}
{\widetilde{\calN_X}^{1+\epsilon}}\right).$$
We have seen in Subsection \ref{subsec:remaining_tensors} that 
$\forall~(i,j,k)\in S'_8, \calN_{ijk,X}=\calN^*_{ijk,X}$. Let us rewrite
$$\widetilde{\calN_X}=\left(\prod\limits_{(i,j,k)\in S'_8} \calN_{ijk,X}\right)
\left(\prod\limits_{(i,j,k)\in \{(2,3,3),(3,2,3),(3,3,2)\}} \calN_{ijk,X}\right)$$
and
$$\widetilde{\calN_X^*}=\left(\prod\limits_{(i,j,k)\in S'_8} \calN^*_{ijk,X}\right)
\left(\prod\limits_{(i,j,k)\in \{(2,3,3),(3,2,3),(3,3,2)\}} \calN^*_{ijk,X}\right).$$
We have seen in Section \ref{subsec:T233} that by imposing the constraints $\eqref{eqn:xi2}$
we have $\calN_{233,X}=\calO(N^2\calN_{233,X}^*)$, $\calN_{323,X}=\calO(N^2\calN_{323,X}^*)$ and $\calN_{332,X}=\calO(N^2\calN_{332,X}^*)$.
Therefore we have 
$$\prod\limits_{(i,j,k)\in \{(2,3,3),(3,2,3),(3,3,2)\}} \calN_{ijk,X}
=\calO\left(N^6\prod\limits_{(i,j,k)\in \{(2,3,3),(3,2,3),(3,3,2)\}} \calN^*_{ijk,X}\right),$$
and thus
$\widetilde{\calN_X} = \calO(N^6\widetilde{\calN_X^*})$ and the number of terms in the direct sum that we obtain is 
$$\Omega\left(\dfrac{\widetilde{\calT_X}}
{N^{6(1+\epsilon)}\widetilde{\calN^*_X}^{\epsilon}}\right).$$
For any $(i,j,k)\in \overline{S}_8$, by definition we have
$$\calN^*_{ijk,X} \le |\overline{S}_{ijk}|^{a(ijk)N} \le |S_4|^{a(ijk)N} = \left(\sum\limits_{l=1}^5 l \right)^{a(ijk)N} = 15^{a(ijk)N}$$ and thus 
$\widetilde{\calN^*_X} \le 15^N$, and the number of terms in the direct sum is 
$$\Omega\left(\dfrac{\widetilde{\calT_X}}
{15^{N\epsilon} N^{6(1+\epsilon)}}\right).$$
Replacing $\widetilde{\calT_X}$ by its expression, this is equal to 
$$r_2=\Omega\left(\dfrac{1}{15^{N\epsilon} N^{6\epsilon + 6 + \sum\limits_{(i,j,k)\in \overline{S}_8}\frac{i}{2}}}
\left[\prod\limits_{(i,j,k)\in \overline{S}_8}\left(\dfrac{1}{\prod\limits_{u=0}^i A_{ijk}(u)^{A_{ijk}(u)}}\right)^{a(ijk)}
\right]^N\right).$$
We summarize the result of this section in the following theorem.

\begin{theorem}\label{th:extra2}
Let $q$ be any positive integer.
Let $a$ be any function from $\Pp(S_8)$ satisfying the symmetry constraint $\eqref{eqn:symm1}$.
For all $(i,j,k)\in \overline{S}_8$, let $a_{ijk}$ be functions from $\Pp(\overline{S}_{ijk})$ 
satisfying the constraints
 $\eqref{eqn:symm2}$, $\eqref{eqn:xi2}$ and $\eqref{eqn:D3}$. 
Then for any $\epsilon > 0$, the trilinear form $\bigotimes\limits_{(i,j,k)\in \overline{S}_8} T_{ijk}^{\otimes a(ijk)N}$ admits a restriction which is a direct sum of $r_2$ tensors, all isomorphic to
$$\bigotimes\limits_{(u,v,w)\in \overline{S}_8}\left(\bigotimes\limits_{(i,j,k) \in \overline{S}_{uvw}}
V_{uvw}(ijk)^{\otimes a_{uvw}(ijk)a(uvw)N}\right).$$
\end{theorem}

\section{Third extraction}\label{sec:extra3}

\begin{sloppypar}From the second extraction of Section \ref{sec:extra2} appeared tensors of the form $V_{uvw}(ijk)$, which are formed from tensors $T_{ijk}, (i,j,k) \in S_4$. We will see in Section \ref{sec:full} that all the tensors $T_{ijk}$ with 
${(i,j,k)\in S_4 \setminus \overline{S}_4}$, where ${\overline{S}_4=\{(2,1,1),(1,1,2),(2,1,2)\}}$, correspond to matrix products. The last extraction, that we now realize, deals with the tensors $T_{211}$, $T_{121}$ and $T_{112}.$\end{sloppypar}

\subsection{Properties of the tensors {\boldmath$T_{211}$}, {\boldmath$T_{121}$} and {\boldmath$T_{112}$}}\label{subsec:prop211}

We first focus on the tensor $T_{211}$ and analyze it as done in Ref.~\cite{LeGallFOCS12}. This tensor can be written as $T_{211}=t_{011}+t_{101}+t_{110}+t_{200}$, where
\begin{eqnarray*}
t_{011}&=&\sum_{i=1}^q x_{0,q+1}y_{i,0}z_{i,0},\\
t_{101}&=&\sum_{i,k=1}^q x_{i,k}y_{0,k}z_{i,0},\\
t_{110}&=&\sum_{i,k=1}^q x_{i,k}y_{i,0}z_{0,k},\\
t_{200}&=&\sum_{k=1}^q x_{q+1,0}y_{0,k}z_{0,k}.
\end{eqnarray*}
For $(I,J,K) = (I_l,J_l,K_l)_{l\in\Ni{1}{n}} \in S_2^{n}$, define $t_{IJK} = \bigotimes\limits_{l\in\Ni{1}{l}} t_{I_l J_l K_l}$.

We now describe an extraction from 
\[
T_{211}^{\otimes 2m} = \sum\limits_{(I,J,K) \in S_2^{2m}} t_{IJK},
\]
 where $m$ is an integer.
Note that for $(i,j,k),(i',j',k') \in S_2$, 
if $i\neq i'$ the $x$ variables in $t_{ijk}$ and $t_{i'j'k'}$ are distinct,
if $j\neq j'$ the $y$ variables in $t_{ijk}$ and $t_{i'j'k'}$ are distinct, and
if $k\neq k'$ the $z$ variables in $t_{ijk}$ and $t_{i'j'k'}$ are distinct. 
Thus, for $(I,J,K),(I',J',K') \in S_2^{2m}$, 
if $I\neq I'$ the $x$ variables in $t_{IJK}$ and $t_{I'J'K'}$ are distinct,
if $J\neq J'$ the $y$ variables in $t_{IJK}$ and $t_{I'J'K'}$ are distinct, and
if $K\neq K'$ the $z$ variables in $t_{IJK}$ and $t_{I'J'K'}$ are distinct. 
Fix $b\in\,]0,1[\,\cap\,\Q$ and assume ($m$ will later go to infinity) that $bm\in\N$.
Define $ a_{211}(011)= a_{211}(200) = (1-b)/2$, $a_{211}(101)=a_{211}(110)=b/2$, with projections $A_{211},B_{211},C_{211}$. Note that by definition of $a_{211}$, $B_{211}=C_{211}$.
We set to $0$ all the $x$ variables but the ones which appear in a $t_{IJK}$ with $I$ of type $A_{211}$, and all the $y$ (resp. $z$) except the ones which appear in a $t_{IJK}$ with $J$ (resp. $K$) of type $B_{211}$.

It was shown in Subsection 6.1 of \cite{LeGallFOCS12} that this leads to a sum of forms $t_{IJK}$ isomorphic to 
$$t_{011}^{\otimes (1-b)m}\otimes t_{101}^{\otimes bm}\otimes t_{110}^{\otimes bm}\otimes t_{200}^{\otimes (1-b)m}
\cong \braket{q^{2bm},q^{2bm},q^{2(1-b)m}}$$
and that, adopting the same notations as in the previous extractions of Sections \ref{sec:extra1} and \ref{sec:extra2}, we have 
\begin{eqnarray*}
\calT_{211,X}&=&{2m \choose (1-b)m,(1-b)m,2mb}=\Theta\left(\frac{1}{m}\cdot \left[\frac{2}{(2b)^b(1-b)^{1-b}}\right]^{2m}\right)\\
\Nn_{211,X}=\Nn_{211,X}^*&=&{2mb \choose mb}=\Theta\left(\frac{1}{\sqrt{m}}\cdot  \left[2^b\right]^{2m}\right)\\
\calT_{211,Y}=\calT_{211,Z}&=&{2m \choose m}=\Theta\left(\frac{1}{\sqrt{m}}\cdot  \left[2\right]^{2m}\right)\\
\Nn_{211,Y}=\Nn_{211,Z}=\Nn_{211,Y}^*=\Nn_{211,Z}^*&=&{m \choose m(1-b)}{m \choose m(1-b)}=\Theta\left(\frac{1}{m}\cdot  \left[\frac{1}{b^b(1-b)^{1-b}}\right]^{2m}\right) .\\
\end{eqnarray*}

The forms $T_{112}$ and $T_{121}$ can be analyzed in the same way as $T_{211}$ by permuting the roles of the $x$ variables, the $y$ variables and the $z$ variables.

\subsection{Joint extraction of the tensors {\boldmath$T_{211}$}, {\boldmath$T_{121}$} and {\boldmath$T_{112}$}}\label{sub:joint}

From the extraction of Sections \ref{sec:extra1} and \ref{sec:extra2}, we obtain tensors isomorphic to 
$$T_{112}^{\otimes \alpha_{112}N}\otimes T_{121}^{\otimes \alpha_{121}N}\otimes T_{211}^{\otimes \alpha_{211}N}$$
where for $(i,j,k)\in\{(2,1,1),(1,1,2),(2,1,2)\}$, 
\begin{align*}
\alpha_{ijk} = &\sum\limits_{(u,v,w) \in\overline{S}_8, \exists (i',j',k') \in S_4, ((i,j,k),(i',j',k')) \in S_{uvw}}
a(uvw)a_{uvw}(ijk) ~+ \\
&\sum\limits_{(u,v,w) \in \overline{S}_8, \exists (i',j',k') \in S_4, ((i',j',k'),(i,j,k)) \in S_{uvw}}
a(uvw)a_{uvw}(i'j'k').
\end{align*}
Note that the symmetry conditions $\eqref{eqn:symm1}$ and $\eqref{eqn:symm2}$ imply that $\alpha_{112}=\alpha_{121}$.

\begin{sloppypar}Using the parameters from Subsection \ref{subsec:prop211}, we realize a joint extraction on the tensor ${T_{112}^{\otimes \alpha_{112}N}\otimes T_{121}^{\otimes \alpha_{112}N}\otimes T_{211}^{\otimes \alpha_{211}N}}$, just as in Subsection \ref{subsec:joint}.
We actually use a constant $b\in\,]0,1[$ for defining the types used for $T_{112}$ and $T_{121}$ and another constant $\tilde{b}$ for the type used for $T_{211}$. We impose the constraint\end{sloppypar}
\begin{equation}
\tag{E3}
\dfrac{\left(2^{\tilde{b}}\right)^{\alpha_{211}}}{\left(b^b (1-b)^{1-b}\right)^{\alpha_{112}}} \ge
\dfrac{\left(2^b\right)^{\alpha_{112}}}{\left(\tilde{b}^{\tilde{b}} (1-{\tilde{b}})^{1-\tilde{b}}\right)^{\alpha_{211}}}
\label{eqn:E3}
\end{equation}
which ensures that 
$\calN_{211,Y}\calN_{112,Y}\calN_{121,Y} = 
\calN_{211,Z}\calN_{112,Z}\calN_{121,Z} = \calO\left(\calN_{211,X}\calN_{112,X}\calN_{121,X}\right)$.
We then get a direct sum of 
$$\Omega\left(\dfrac{\calT_{211,X}\calT_{112,X}\calT_{121,X}}
{(\calN_{211,X}\calN_{112,X}\calN_{121,X})^\epsilon}\right)$$ trilinear forms, all isomorphic to 
$$\widehat{T_{211}}=\braket{q^{(\alpha_{112}+ \alpha_{211}\tilde b)N},q^{(\alpha_{112}+ \alpha_{211}\tilde b)N},q^{(2\alpha_{112}b+\alpha_{211}(1-\tilde b))N}}.$$
Replacing the $\calT$ by their values, and using the bounds 
$\calN_{211,X} \le 4^{\alpha_{211}N}$, $\calN_{112,X}=\calN_{121,X} \le 4^{\alpha_{112}N}$, we get that the number of terms in the direct sum is
$$r_3=\Omega\left(\dfrac{1}{4^{(\alpha_{211}+2\alpha_{112})N\epsilon}N^2}\left[\dfrac{2^{2\alpha_{112} + \alpha_{211}}}
{\left((2\tilde{b})^{\tilde{b}}(1-\tilde{b})^{1-\tilde{b}}\right)^{\alpha_{211}}}\right]^N\right).$$
We summarize the result of this last extraction in the following theorem.

\begin{theorem}
For any positive $\alpha_{211},\alpha_{112}$, for any $b,\tilde{b}\in\,]0,1[\,\cap\,\Q$ satisfying $\eqref{eqn:E3}$ 
and for any $\epsilon>0$, the trilinear form
$$T_{112}^{\otimes \alpha_{112}N}\otimes T_{121}^{\otimes \alpha_{112}N}\otimes T_{211}^{\otimes \alpha_{211}N}$$
admits a restriction which is a direct sum of $r_3$ trilinear forms, all isomorphic to $\widehat{T_{211}}$.
\end{theorem}
\section{The full extraction}\label{sec:full}

Let us first extend the definition of the values $\alpha_{ijk}$, which were introduced only for $(i,j,k)\in \overline{S}_4$ in Section \ref{sub:joint}, to all triples in $S_4$:
for any $(i,j,k)\in S_4$ define
\begin{align*}
\alpha_{ijk} = &\sum\limits_{(u,v,w) \in \overline{S}_8, \exists (i',j',k') \in S_4, ((i,j,k),(i',j',k')) \in S_{uvw}}
a(uvw)a_{uvw}(ijk) N ~+ \\
&\sum\limits_{(u,v,w) \in \overline{S}_8, \exists (i',j',k') \in S_4, ((i',j',k'),(i,j,k)) \in S_{uvw}}
a(uvw)a_{uvw}(i'j'k') N.
\end{align*}
From the three consecutive extractions described in Sections \ref{sec:extra1}--\ref{sec:extra3}, we get a direct sum of $r_1 r_2 r_3$ trilinear forms, each of them being isomorphic to 
\begin{equation}
\tag{$\ast$}
\left(\bigotimes\limits_{(i,j,k)\in S_8 \setminus \overline{S}_8} T_{ijk}^{a(ijk)N}\right)
\otimes\left(\bigotimes\limits_{(i,j,k)\in S_4 \setminus \overline{S}_4} T_{ijk}^{\alpha_{ijk} N}\right) 
\otimes\widehat{T_{211}}.
\label{eqn:ast}
\end{equation}

For any $(i,j,k) \in S_4 \setminus \overline{S}_4$, the trilinear form $T_{ijk}$ represent a matrix product (cf. \cite{Coppersmith+90}):
\begin{align*}
T_{004}\cong T_{040}\cong T_{400}\cong~&\langle 1,1,1\rangle\\
T_{013}\cong T_{031}\cong~&\langle 1,1,2q\rangle\\
T_{103}\cong T_{301}\cong~& \langle 2q,1,1\rangle\\
T_{130}\cong T_{310}\cong~& \langle 1,2q,1\rangle\\
T_{022}\cong~&\langle 1,1,q^2+2\rangle\\
T_{202}\cong~&\langle q^2+2,1,1\rangle\\
T_{220}\cong~&\langle 1,q^2+2,1\rangle .
\end{align*}
It can be seen from the definitions of the trilinear form $T_{ijk}$, $(i,j,k)\in S_8 \setminus \overline{S}_8$ that they also  represent a matrix product. We have:
\begin{align*}
T_{008}\cong T_{080}\cong T_{800}\cong~&\braket{1,1,1}\\
T_{017}\cong T_{071}\cong~&\braket{1,1,4q}\\
T_{107}\cong T_{170}\cong~&\braket{4q,1,1}\\
T_{701}\cong T_{710}\cong~&\braket{1,4q,1}\\
T_{026}\cong T_{062}\cong~&\braket{1,1,6q^2 + 4}\\
T_{206}\cong T_{260}\cong~&\braket{6q^2 + 4,1,1}\\
T_{602}\cong T_{620}\cong~&\braket{1,6q^2 + 4,1}\\
T_{035}\cong T_{053}\cong~&\braket{1,1,4q^3 + 12q}\\
T_{305}\cong T_{350}\cong~&\braket{4q^3 + 12q,1,1}\\
T_{503}\cong T_{530}\cong~&\braket{1,4q^3 + 12q,1}\\
T_{044}\cong~&\braket{1,1,q^4 + 12q^2 + 6}\\
T_{404}\cong~&\braket{q^4 + 12q^2 + 6,1,1}\\
T_{440}\cong~&\braket{1,q^4 + 12q^2 + 6,1}.
\end{align*}
As we have seen in Section \ref{sec:extra3}, $\widehat{T_{211}}$ is also a matrix product. Hence, the $r_1 r_2 r_3$ isomorphic trilinear forms that we have extracted all represent the same matrix product. By the symmetry constraints \eqref{eqn:symm1} and \eqref{eqn:symm2}, we get that this matrix product is of the form 
$\braket{Q^N,Q^N,R^N}$. The expressions of $Q$ and $R$ are obtained by replacing in \eqref{eqn:ast} the $T_{ijk}$ and $\widehat{T_{211}}$ by the matrix products they correspond to. We refrain from giving here the complete expressions for $R$ and $Q$ since the formulas are extremely long (they can be found in the files of the programs used for the numerical analysis \cite{fileurl}).
We have shown:
$$r_1 r_2 r_3 \cdot \braket{Q^N,Q^N,R^N} \le F^{\otimes N}.$$
As we already saw in Subsection \ref{subsec:trilin}, $\underline{R}\left(F\right)\le (q+2)^4$ and thus by submultiplicativity of the border rank
$\underline{R}\left(F^{\otimes N}\right)\le (q+2)^{4N}$.
By Sch\"onhage's asymptotic sum inequality (Theorem~\ref{thm:Schonhage}), we have:
$$r_1 r_2 r_3 Q^{N\omega(\frac{\log R}{\log Q})}\le (q+2)^{4N},$$ and taking the $N$-th root, we get:
$$(r_1 r_2 r_3)^{\frac{1}{N}} Q^{\omega(\frac{\log R}{\log Q})}\le (q+2)^4.$$
Let 
$$\calM = \lim\limits_{\epsilon \rightarrow 0} \lim\limits_{N \rightarrow \infty} (r_1 r_2 r_3)^{\frac{1}{N}}
=\dfrac{1}{\prod\limits_{i=0}^8 A(i)^{A(i)}} 
\prod\limits_{(i,j,k)\in \overline{S}_8}\left(\dfrac{1}{\prod\limits_{u=0}^i A_{ijk}(u)^{A_{ijk}(u)}}\right)^{a(ijk)}
\dfrac{2^{2\alpha_{112} + \alpha_{211}}}{\left((2\tilde{b})^{\tilde{b}}(1-\tilde{b})^{1-\tilde{b}}\right)^{\alpha_{211}}}.$$
We have
$$\calM Q^{\omega(\frac{\log R}{\log Q})}\le (q+2)^4.$$

We summarize the result of the whole process in the following theorem. 

\begin{theorem}\label{thm:main}
Let $q$ be any positive integer. Let $a$ be any function from $\Pp(S_8)$ satisfying the constraints $\eqref{eqn:symm1}$, $\eqref{eqn:xi}$ and $\eqref{eqn:C3}$.
For all $(u,v,w)\in \overline{S}_8$, let $a_{uvw}$ be functions from $\Pp(\overline{S}_{uvw})$ 
satisfying the constraints $\eqref{eqn:symm2}$, $\eqref{eqn:xi2}$ and $\eqref{eqn:D3}$. 
Finally, let $b$ and $\tilde{b}$ be any two values from $]0,1[\,\cap\,\Q$ satisfying the constraint $\eqref{eqn:E3}$. Then the following inequality holds:
$$\calM Q^{\omega(\frac{\log R}{\log Q})}\le (q+2)^4.$$
\end{theorem}

Theorem \ref{thm:main} enables us to obtain our new upper bounds on $\omega(k)$: for any $k$,
if we find values $q$, $a$, $a_{uvw}$ for each $(u,v,w)\in \overline{S}_8$, $b$ and $\tilde{b}$ satisfying the constraints in Theorem \ref{thm:main} such that $\frac{\log R}{\log Q} = k$ and $\calM Q^\nu \ge (q+2)^4$ for some $\nu$ , then we get, since $\calM Q^{\omega(k)} \le (q+2)^4$, that $\omega(k) \le \nu$. The bounds given in Table \ref{table_results4} and Theorem~\ref{th:ma} are obtained by finding the optimal values for $q$, $a$, $a_{uvw}$ for each $(u,v,w)\in \overline{S}_8$ , $b$ and $\tilde{b}$ by numerical analysis using Maple. 
The source code of the Maple programs used for the numerical analysis, which include the complete formulas for the terms $R$ and $Q$, is available at~\cite{fileurl}.

\section*{Acknowledgements.}
This work is partially supported by ERC QCC, ANR grant RDAM, JSPS Grant-in-Aid for Young Scientists~(A) No.~16H05853 and JSPS Grant-in-Aids for Scientific Research~(A) No.~15H01677 and 16H01705.
\bibliographystyle{plain}
\bibliography{RMMbib}
\end{document}